\documentclass[journal]{IEEEtran}
\usepackage[edges]{forest}
\usetikzlibrary{arrows.meta}
\usepackage{graphicx,times,amsmath,cite,enumerate}
\usepackage{kantlipsum}
\usepackage{threeparttable}
\usepackage{epsfig}
\usepackage{amssymb}
\usepackage{latexsym}
\usepackage{psfrag}
\usepackage{setspace}
\usepackage{color}
\usepackage{verbatim}
\usepackage{amsthm}
\usepackage{footnote}
\usepackage{textcase}
\usepackage{array}
\newcolumntype{P}[1]{>{\centering\arraybackslash}p{#1}}
\usepackage{float}
  
\usepackage{multirow}
\usepackage{textcomp}
\usepackage{forest}
\PassOptionsToPackage{hyphens}{url}
\usepackage{url}
\usepackage{physics}
\usepackage{hyperref}
\usepackage{etoolbox}
\usepackage[edges]{forest}
\usepackage[ruled,vlined]{algorithm2e}
\usepackage{siunitx}

\usepackage{subfigure}
\usepackage{physics}
\usepackage{tikz}
\usetikzlibrary{automata, positioning}
\usepackage{nomencl}
\DeclareMathOperator*{\argmin}{arg\,min}

\theoremstyle{definition} 
\newtheorem{theorem}{Theorem}
\theoremstyle{remark} 
\newtheorem{rem}{Remark}

\begin{document}
\title{Cyber Insurance Against Cyberattacks \\ on Electric Vehicle Charging Stations}
\author{Samrat Acharya, \emph{Student Member, IEEE}, Robert Mieth,  \emph{Member, IEEE}, Charalambos Konstantinou, \emph{Senior Member, IEEE},~Ramesh Karri, \emph{Fellow, IEEE}, and Yury Dvorkin \emph{Member, IEEE}}

\maketitle
 \begin{abstract}
Cyberattacks in the energy sector are commonplace. Load-altering cyberattacks launched via the manipulations of high-wattage appliances and assets are particularly alarming, as they are not continuously monitored by electric power utilities. Public Electric Vehicle Charging Stations (EVCSs) are among such high-wattage assets. Even EVCSs monitored by the electric power utilities and protected by state-of-the-art defense mechanisms are vulnerable to cyberattacks.
Such cyberattacks cause financial losses to the EVCSs. In this paper, we propose cyber insurance for EVCSs to hedge the economic loss due to such cyberattacks and develop a data-driven cyber insurance design model for public EVCSs. Under mild modeling assumptions, we derive an optimal cyber insurance premium. Then, we ensure the robustness of this optimal premium and investigate the risk of insuring the EVCSs using a suitable risk assessment metric (Conditional Value-at-Risk). A case study with data from EVCSs in Manhattan, New York illustrates our results. Our results demonstrate that risk assessment is crucial for designing insurance premiums. Furthermore, the premium increases in proportion to the loss coverage offered for the EVCSs. This work informs the stakeholders involved in the roll-out and operation of public EVCSs about the benefits of cyber insurance and suggests that insurance premiums can be reduced by deploying state-of-the-art defense
mechanisms.
\end{abstract}

 
\IEEEpeerreviewmaketitle

\section*{Nomenclature}
\addcontentsline{toc}{section}{Nomenclature}
\begin{IEEEdescription}[\IEEEusemathlabelsep\IEEEsetlabelwidth{$i,j,s \in J_0$}]

\item[\textit{Acronyms}]
\item[BL] Bi-Level.
\item[C\&CG] Column-and-Cut Generation. 
\item[CDF] Cumulative Distribution Function.
\item[CVaR] Conditional Value-at-Risk.
\item [DoS] Denial-of-Service
\item[DLMP] Distribution Locational Marginal Price.
\item[EV] Electric Vehicle.
\item[EVCS] Electric Vehicle Charging Station.
\item[LL] Lower Level.
\item[LMP] Locational Marginal Price.
\item[ML] Middle-Level.
\item[SMP] Semi-Markov Process.
\item[TL] Tri-Level.
\item[UL] Upper-Level.
\item[WC] Worst-Case.

\item[\textit{Sets and Indexes}]
\item[$\mathcal{S}$] Set of typical days, indexed by $s$.
\item[$\mathcal{T}$] Set of hours in a day, indexed by $t$.
\item[$\mathcal{B}$] Set of buses, indexed by $b$.
\item[$\mathcal{I}$] Set of generators, indexed by $i$.
\item[$\mathcal{I}_b$] Set of generators connected to bus $b$, $\mathcal{I}_b \in \mathcal{I}$.
\item[$\mathcal{L}$] Set of power lines, indexed by $l$.
\item[l(o), l(r)] Indices of sending, receiving buses of line $l$.  

\item[\textit{Parameters}]
\item[$C_i^g$] Energy price offered by generator $i$, $\$/MWh$.
\item[$\overline{G}_i$] Maximum power output by generator $i$, $MW$.
\item[$\overline{F}_l$] Flow limit on power line $l$, $MW$.
\item[$z_l$] Reactance of line $l$, $\si{\ohm}$.
\item[$d_t^{s}$] EVCS power demand at time $t$ on day $s$, $MW.$
\item[$d^{d,s}_{b,t}$] Nodal power demand other than $d_t^s$ at bus $b$ at time $t$ on day $s$, $MW$.
\item[$r$] Insurance profit loading factor.
\item[$\phi^{s}$] Likelihood of day $s$.
\item [$\gamma$] EVCS cyberattack risk sharing factor.
\item[$\kappa$] Coefficient of increment in insurance premium per cyberattack on EVCS.
\item[$\mathbb{P}(A)$] Probability of a cyberattack on EVCS during the insurance period.
\item[$A_h$] Number of cyberattacks on EVCS in the past.
\item[$\rho$] Penalty to EVCS due to cyberattack, $\$/kW$.

\item[\textit{Variables}]
\item[x] Insurance premium for insurance period.
\item[$\lambda_t^c$] EV charging price offered by an EVCS at time $t$, $\$/kWh$ .
\item[$\lambda_{t,b}^u$] Electricity price offered by a power utility at time $t$ at bus $b$, $\$/kWh$ .
\item[$g_{i,t}^{s}$] Power output of generator $i$ at time $t$ on day $s$, $MW$.
\item[$\theta_{l,t}^{s}$] Voltage phase angle at line end $l$ at time $t$ on typical day $s$, $rad$.
\item[$f_{l,t}^{s}$] Power flow in line $l$ at time $t$ on day $s$, $MW.$

\end{IEEEdescription}

\section{Introduction}
\label{intro}
\IEEEPARstart{S}{mart} grids have become increasingly prone to cyber intrusions. In 2018, the U.S. Department of Homeland Security issued 1211\% more security threats  for entities in  the U.S. energy sector relative to  2010 \cite{energy_monitor}. Similarly, IBM reported a 13\% increase in financial losses from cyber breaches in the energy sector from 2019 to 2020 \cite{ibm}.
While there are initiatives to mitigate such threats, they cannot be  eliminated  due to the cybersecurity terrain shaped by undiscovered vulnerabilities. Concerns arise due to load-altering (demand-side or grid-edge) cyberattacks that compromise operations of high-wattage demand-side appliances and processes (e.g., Electric Vehicles (EVs), demand-side management and microgrid solutions, grid-interactive building systems, etc.)  \cite{ soltan2018blackiot, amini2016dynamic, raman2019manipulating,acharya2021causative, ospina2020feasibility, botena_go}. These cyberattacks leverage many attack points in demand-side information and communication technologies, including customer devices with poor security hygiene and backdoors in complex supply chains of grid-edge equipment \cite{acharya2020cybersecurity,metere2021securing, sandia}. Furthermore, the demand-side appliances and grid-edge equipment are not directly monitored by the utilities, preventing continuous validation of  trustworthiness. As a result, there is no means to ensure robustness of the appliances against cyberattacks and  to promote cybersecurity compliance and hygiene among electricity consumers. To bridge this gap, this paper designs cyber insurance as a mechanism to promote cybersecurity in smart grids by  hedging the financial risks imposed by load-altering attacks.  The proposed cyber insurance mechanism is compatible with real-world data. As a case study, we demonstrate its effectiveness for Electric Vehicle Charging Stations (EVCSs) using data from Manhattan, New York (NY).

Cyber insurance is designed to cover losses and liabilities caused by cyber incidents such as, data breaches, business interruption, health repercussions and equipment damage \cite{cyberinsurance_def}.
The global cyber insurance market is expected to grow from \$4.8 billion in 2018 to \$28.6 billion in 2026, a compound average annual growth rate of 24.9\% \cite{insurance_market}. The contribution of the energy sector to this market is minuscule relative to the banking and financial services, information technology, and healthcare sectors. However, with the proliferation of high-impact cyberattacks in the energy sector, such as the  Colonial Pipeline attack in May 2021, cyber insurance has gained prominent attention \cite{colonial}. Nevertheless,  present risk modeling and insurance evaluation techniques treat the energy sector similarly to the information technology sector and do not account for the physical aspects of energy delivery \cite{insurance_market}. Such considerations may lead to erroneous insurance policies reducing their viability. A lack of domain-specific risk modeling techniques impact the bottom line of the cyber insurance industry. For example, the loss ratio of the US cyber insurers increased from 47\% in 2019 to 73\% in 2020 \cite{fitch}.

Recent studies in \cite{romanosky2019content, woods2019county} analyze methods for designing cyber insurance premiums. Romanosky \textit{et al.} \cite{romanosky2019content} study 235 cyber insurance policies underwritten between 2007 and 2017 in New York, Pennsylvania, and California and summarize the core insurance design principles and the design parameters (e.g., how to characterize cyber incidents, differentiate risks between customers, and price these risks). We present these in  Section~\ref{sec:loss_model}. Woods \textit{et al.} \cite{woods2019county} study real-world data from 26 cyber insurers in California  and summarize the premium differences across insurance characteristics (e.g.,  coverage ratio, risk exposure,  revenue  of the insured entity). The results from \cite{woods2019county} demonstrate that  risk exposure from cyber incidents are best modeled using Gamma, Log-Normal, Polynomial, Pareto, and Weibull distributions. This departs from the exponential distributions used to deal with adversarial uncertainties in smart grids (e.g., unplanned outages \cite{allan2013reliability}).  From a theoretical perspective, Zhang \textit{et al.}  \cite{zhang2017bi} propose a non-cooperative game between the insurer and the insured to design the optimal premium, where the upper-level models an insurer and the lower-level models the insured entity, which play a zero-sum game with an attacker. Results in  \cite{zhang2017bi} demonstrate that cyber insurance products encourage insured entities to adopt security mechanisms to reduce the success rate of cyberattacks and the insurance premiums. The studies in \cite{romanosky2019content, woods2019county,zhang2017bi} consider generic cyberattack risks and lack a domain-specific context.  This paper focuses on designing a data-driven  optimization for computing cyber insurance premiums informed by insights from real-world  operations of electric power distribution systems with EVCSs, which are a potent load-altering attack vector \cite{DOE, khan2019impact, basnet2021exploring, acharya2020cybersecurity, acharya2020public}. 

Recent studies in \cite{maynard2015business, lau2020cybersecurity,  lau2021coalitional, liu2020actuarial, yang2020premium} consider  cyber insurance mechanisms for the power grids, and \cite{ niyato2017cyber, hoang2017charging} specifically focus on EVs. Maynard and Beecroft \cite{maynard2015business} study the need for cyber insurances in the U.S. power grids using the \textit{Erebos} trojan attack, which can incur financial loss up to \$223.8 billion. Lau \textit{et al.}  \cite{lau2020cybersecurity} show that insurance premiums for transmission companies depend on the allocation of defenses across the network and can be reduced up to $\approx 50\%$, if optimized with domain-specific insights from transmission network operations. Furthermore, Lau \textit{et al.} \cite{lau2021coalitional} show that insurance premiums can be cheaper than in  \cite{lau2020cybersecurity}, if the transmission companies collectively insure themselves without a third-party insurer. Similarly, Liu \textit{et al.} \cite{liu2020actuarial} show that the insurance premiums for the transmission companies increases if they have operational interdependence. This premium increase is caused by the increment in interdependent cyber risks. In \cite{lau2020cybersecurity, liu2020actuarial}, power grid cyberattacks are  modeled using a Semi-Markov Process (SMP), which makes it possible to realistically capture stochastic transitions between ``nominal'' and ``attacked'' system states.  The case studies in \cite{lau2020cybersecurity,lau2021coalitional, liu2020actuarial} use the synthetic IEEE Reliability Test System-96 benchmark. Similarly, using another artificial IEEE benchmark, Yang \textit{et al.} \cite{yang2020premium} demonstrate the effects of  generator ramping rates, bus loading, and  intrusion probabilities on  cyber insurance premiums. The results in \cite{yang2020premium} suggest that  premiums  increase as the likelihood of intrusion, which nonlinearly depends on bus loading, increases. 
Niyato \textit{et al.} \cite{niyato2017cyber} demonstrate that a cyber-insured EV minimizes its cost even during DoS attacks on EVCSs. Furthermore, Honag \textit{et al.} \cite{hoang2017charging} demonstrate that  cyber-insured EV charging is also profit-maximizing when supplying  vehicle-to-grid services to the power grid.
However, the studies in \cite{maynard2015business, lau2020cybersecurity,  lau2021coalitional, liu2020actuarial, yang2020premium, hoang2017charging,niyato2017cyber} are limited in two notable aspects. First, they assume that the insurer  knows  insurance policy parameters, e.g., the probability of an attack on the insured entity, with perfect accuracy. Second,  attack scenarios (e.g., \textit{Erebos} trojan in \cite{maynard2015business}) or attack implementations deal with artificial data sets, which are customized to show profound attack effects.  

Departing from \cite{zhang2017bi, romanosky2019content, woods2019county, maynard2015business, lau2020cybersecurity,  lau2021coalitional, liu2020actuarial, yang2020premium, hoang2017charging,niyato2017cyber}, this paper develops a cyber insurance mechanism that focuses on a realistic load-altering attack vector stemming from electric vehicle charging and uses real-world charging and power grid data from Manhattan, NY. The paper makes four contributions:
\begin{enumerate}
    \item It develops an optimization framework to design an  insurance premium with \textit{(i)} a predetermined (regulated) electricity tariff  mirroring the current practice,  and \textit{(ii)} a dynamic electricity tariff settled using the Distribution Locational Marginal Price (DLMP) framework \cite{caramanis2016co} representing electric power distribution of the future. The first case uses a Bi-Level (BL) game between the insurer and the EVCS. The second one uses a Tri-Level (TL) game between the insurer, EVCS, and distribution system.
    \item In the case of the predetermined tariff, it derives an analytical solution determining the optimal insurance premium  and analyzes its dependence on external factors, i.e., EVCS attack probability, EVCS risk-attitude, and insurance profit loading factor. Under the DLMP tariff, it obtains a numerical solution using a computationally tractable  column-and-cut generation (C\&CG) algorithm.
    \item In both cases, it employs a SMP informed by real-world data to estimate the probability of attack on EVCSs. This  can accurately parameterize the insurance decisions. Further, the premium optimization is made robust using Conditional-Value-At-Risk (CVaR), which internalizes a risk attitude of the EVCS.
    \item It uses EVCS data from Manhattan, NY to study the role of cyber insurance on financial impact of cyberattacks. 
\end{enumerate}

\section{Cyber Insurance Design Principles}
\label{sec:loss_model}
This section describes foundational principles of cyber insurance design as they apply to EVCSs and electric power distribution systems. Consider an EVCS buying a cyber insurance for a time period. The insurer collects a monthly premium payment from the EVCS. In return, the insurer fully or partially covers the EVCS financial loss caused by a cyberattack on the EVCS. This loss can be first-party and third-party. First-party losses refer to the loss directly incurred by the insured entity, e.g., from damage caused by cyberattacks on physical and cyber assets of the EVCS and/or a service interruption. Third-party losses, or liability, refer to payouts and legal costs that the insured entity needs to pay if it is found legally liable. For example, an EVCS may be liable for damages or injuries of its employees, customers, EVs and the power grid caused by a cyberattack. 

One of the key challenges of the optimal cyber insurance design is to assess and evaluate uncertain first- and third-party losses with high accuracy. Both the insurer and insured entity benefit from increasing the accuracy of loss assessments and, therefore, often enter  proprietary, non-public agreements defining various aspects of loss coverage. As a result, they obfuscate the methods for assessing attack risks and optimal insurance premiums. However, regulatory authorities require  that insurers  submit their policies for review \cite{romanosky2019content, deloitte_regulators}. For example, U.S.  insurers  file their issued policies with the National Association of Insurance Commissioners (NAIC), a non-for-profit organization supervising insurance policies \cite{naic}. These authorities check if the insurance policies are: \textit{(i)} affordable to the insured entity, \textit{(ii)} adequate for the insurers to run their business, and \textit{(iii)} uniformly priced across entities with similar risk profiles. 
Following \cite{romanosky2019content}, we summarize three types of insurance policies:
\begin{enumerate}
\item \emph{Flat rate}: The insurer estimates the cyberattack risk of the insured entity and the expected financial loss for a given time period. Then, the insurer computes the flat rate premium to trade-off its desired profit  and  the risk exposure levels of the insured entity. This premium policy can be applied uniformly for similar small businesses.  
    \item \emph{Flat rate with hazard groups}: In addition to the flat rate policy above, this policy labels cyberattack risks of the insured entities into low, medium, and high risks. Such risks are qualitatively analyzed based on the sensitivity of the cyber space of the insured entities, i.e., the volume of their networks and connected devices. Each hazard group is assigned its own flat-rate policy. 
    \item \emph{Base rate}: In addition to the cybersecurity factors described above, this policy accounts for the asset value or revenue of the insured entity and industry-dependent policy factors (e.g., cyberattack frequency and  historical insurance claims). Furthermore, the base rate calculation often incorporates the use of the security mechanisms of the insured entity to reduce cyberattack risks (e.g., multi-factor authentications and data encryption).

\end{enumerate}
In this paper, we use the base rate premium design principle to assess the cyber insurance of the EVCSs. In contrast to the two flat-rate policies, the base-rate policy allows the insurer to incorporate the risk-attitude of the insured entity (e.g., the portion of loss coverage, attack history, cyber interfaces, and  defenses) and data-driven design of the insurance policy. The cyber interfaces of the EVCS with external entities, such as the power grid, EV fleet, etc, inform the insurer about security threats stemming from their vulnerabilities (e.g., \cite{hodge2019vehicle,soykan2021disrupting,acharya2020public}). The insurers use EVCS security recommendations (e.g., \cite{encs}) to assess the overall security trustworthiness of EVCSs.
 
\section{Insurance Design Models}
\label{sec:modesl}
\subsection{Predetermined Tariff}
\label{sec:bilevel_equiv_insurance_premium}

\begin{figure}[!t]
\centering

\subfigure[\label{fig:bi-level-model}]{
\includegraphics[width=0.48\columnwidth, clip=true, trim= 0mm 0mm 0mm 0mm]{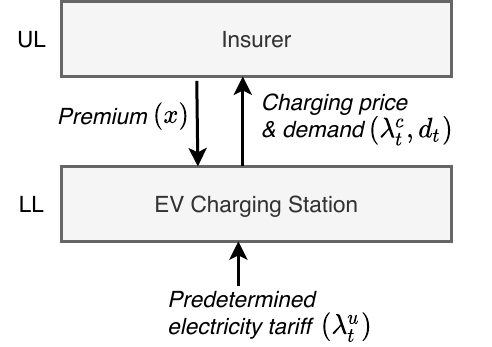}}
\hfill
\subfigure[\label{fig:tri-level problem}]{
\includegraphics[width=0.48\columnwidth, clip=true, trim= 0mm 0mm 0mm 0mm]{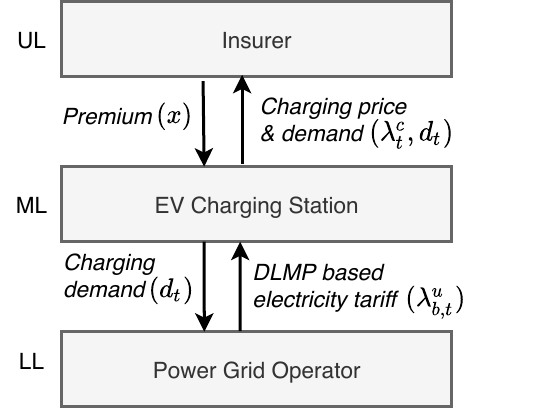}}
\caption{Structure of optimization models for  EVCS cyber insurance: (a) Bi-Level (BL) for the predetermined electricity tariff and (b) Tri-Level (TL) for the DLMP based electricity tariff.}
\end{figure}

Many EVCSs operate in power grids with a predetermined (regulated) electricity tariff. For example, Con Edison of New York established a customized volumetric tariff for EV charging \cite{con_ED}. Under this tariff, interactions between the insurer and EVCS can be modeled as a non-cooperative game structure  in Fig.~\ref{fig:bi-level-model}, where the Upper-Level (UL) models the insurer and the Lower-Level (LL) is the insured EVCS. This game can be expressed as a BL optimization: 
\allowdisplaybreaks
\begin{subequations}
\label{eq:ul} 
\begin{align} \label{eq:objective_upper_bi} 
    \min_{\Xi^{I}}&  ~ x  \\ \label{eq:loss_revenue}
     s.t.&~ CL(d,\lambda_t^{c,*}) = \underbrace{\frac{\gamma (1+\kappa A_{h})\mathbb{P}(A)}{1-r} }_{\text{Policy factors $\Pi$}}\underbrace{\sum_{s=1}^{S}\phi^{s}\sum_{t=1} ^{T}d_{t}^{s}\lambda_t^{c,*}}_{\text{EVCS revenue}}, \\  \label{eq:premium_bound}
    & x \geq CL,
    \end{align}
\end{subequations}
\begin{subequations}
\label{eq:ml}
\begin{align}
\label{eq:objective_lower_bi}
    \hspace{6mm} &\boldsymbol{\lambda^{c,*}}  = \argmin_{\Xi^{E}}~ ||{\boldsymbol{\lambda^c}}||_2^2  \\ \notag
     & s.t. ~C(\lambda^c, \lambda_t^{u*}, \hat x^*)\! =\!\overbrace{(1- \mathbb{P}(A))\!\sum_{s=1}^{S}\phi^{s}\!\sum_{t=1}^{T}d_{t}^{s}( \lambda_t^{u,*} -\lambda_t^{c})}^{\text{Net cost during unattacked time}}  \\  \label{eq:EVCS_cost}
    & +\underbrace{\mathbb{P}(A) \sum_{s=1}^{S}\phi^{s}\sum_{t=1}^{T}(d_{t}^{s} ( \rho -\gamma\lambda_t^{c} ))}_{\text{Net cost during attack}} + \underbrace{\sum_{s=1}^{S}\phi^{s}\sum_{t=1}^{T} \hat x^{*} d_t^s}_{\text{Cost of insurance}} \leq 0.   
    \end{align}
\end{subequations}
where the insurer and EVCS are  modeled  in Eqs.~\eqref{eq:objective_upper_bi}-\eqref{eq:premium_bound} and   Eqs.~\eqref{eq:objective_lower_bi}-\eqref{eq:EVCS_cost} with decisions  sets ${\Xi^{I}} = \{x \in \mathbb{R}: x\geq 0\}$ and  ${\Xi^{E}} = \{\lambda_t^c \in \mathbb{R}:\lambda_t^c \geq 0\}$, and $\boldsymbol{\lambda^c} = [\lambda_{1}^c, \ldots, \lambda_T^c]^{\top}$. 
\label{sec:up}
The insurer in Eq.~\eqref{eq:ul} seeks to determine the lowest insurance premium $(x)$ for a given period that  recovers the expected cyber loss $(CL)$ caused by a cyberattack on the EVCS. Computing the minimal premium based on the data available to the insurer, its risk attitude and choice of insurance design parameters will ensure the competitiveness of the offered policies. Additionally,  some insurance regulatory authorities (e.g., NAIC \cite{naic} and International Association of Insurance Supervisors (IAIS) \cite{iais}) require the insurers to charge premiums that are affordable to their insured entities, which can be measured based on the revenue of insured entities as described in Section~\ref{sec:loss_model}. The expected loss in Eq.~\eqref{eq:loss_revenue} is computed as  the  product of the expected EVCS revenue and industry-dependent policy factors $\Pi =\{ \mathbb{P}(A), r, \gamma, \kappa, A_h\}$. The expected revenue is computed over a given set of typical operational days (indexed by $s \in \mathcal{S}$ and with  the  likelihood  of $\phi^s$), which are  provided by  the EVCS to the insurer. Policy factors $\Pi$ are computed by the insurer and include an attack probability during the insured period $(\mathbb{P}(A))$, a profit loading factor $(r)$,  a risk sharing factor $(\gamma\in[0,1])$, a premium increment factor for historical cyber incidents $(\kappa)$, and a count of historical cyber incidents on the EVCS $(A_h)$. Parameter $\mathbb{P}(A)$ captures the cyberattack risk of an EVCS. Accordingly, a greater  value of $r$ implies a greater confidence level in estimating the  cyber losses and, thus, potentially, lower profit margins of the insurer, and vice versa. Moreover, the insurer can use $r$ to customize the premium for EVCSs. Parameter $r$ can be set lower for an EVCS using defense mechanisms as compared to an EVCS which is not using any defense mechanisms.
The risk sharing factor makes it possible to arbitrage risk sharing between the insurer and EVCS, where $\gamma=1$ indicates a complete transfer of the cyber risk to the insurer and  $\gamma=0$ indicates no risk sharing, i.e., no insurance.
Parameter $\kappa \geq  0$ imposes a  penalty for $A_h$, while  $\kappa=0$ indicates no marginal increment for $A_h$. Notably, parameters $r$, $\kappa$, and $\gamma$ differentiate the base rate policy from the flat rate policies described in Section~\ref{sec:loss_model} by introducing additional degrees of flexibility to match characteristics of specific insured parties more accurately.

\label{sec:EVCS_bi}

In turn, the EVCS in Eq.~\eqref{eq:ml} minimizes the hourly EV charging rate $(\lambda_t^c)$ for $t \in \mathcal{T}$ hours that recovers the expected EVCS operating cost $(C)$ during the insured period. The EVCS minimizes the EV charging price to compete with its peer EVCSs. The expected operating cost in Eq.~\eqref{eq:EVCS_cost} includes three components:  \textit{(i)} net cost during the unattacked time (probability is $1-\mathbb{P}(A)$), \textit{(ii)} net cost during attacks (probability is $\mathbb{P}(A)$), and \textit{(iii)} insurance premium cost.  During the unattacked period (i.e., the first term in Eq.~\eqref{eq:EVCS_cost}), the EVCS operates normally. Thus, the net EVCS cost is calculated as  the EVCS revenue collected from charging EVs minus the the electricity cost paid to the power utility based on the electricity tariff $(\lambda_t^{u,*})$. During the attack on EVCS (i.e., the second term in Eq.~\eqref{eq:EVCS_cost}), the EVCS does not operate. Thus, the EVCS not only loses revenue from EV charging during the attack period (first-party loss), but also loses the confidence of EV drivers to use the attacked EVCS in the future and may be liable for damages of its EVs, customers, or the power grid (third-party loss). These losses are captured by $\mathbb{P}(A) \sum_s \phi^s \sum_t d_t^s\rho$, where $\rho$ measured in $(\$/kW)$ is the penalty incurred by the attack. However, because the EVCS is insured, it is entitled to receive the insurance payment equal to  $\gamma$ times the lost revenue. Thus, the net cost under attack is the loss in revenue and goodwill minus the insurance claim.
Irrespective of the state of the attack on EVCS, the EVCS pays the insurance premium, as shown in the third term of Eq.~\eqref{eq:EVCS_cost}. Notably, we denote $\sum_{s}\phi ^{s}\sum_{t}\hat x^{*}d_t^{s} = x^{*}$ in Eq.~\eqref{eq:EVCS_cost}.
All  three terms of  Eq.~\eqref{eq:EVCS_cost} are stochastic over $s \in \mathcal{S}$  days with likelihood  $\phi^{s}$.

As Fig.~\ref{fig:bi-level-model} shows, the BL optimization in Eqs.~\eqref{eq:objective_upper_bi}-\eqref{eq:EVCS_cost} requires an exchange of information on EV charging price $\lambda^c$, EV charging demand $d$, and premium $x$ between the EVCS and the insurer. The insurer in Eq.~\eqref{eq:ul} optimizes premium $x$ for the insured period based on $d$ and $\lambda^c$ submitted by the EVCS and presents the premium to the EVCS in Eq.~\eqref{eq:ml}. Then the EVCS optimizes $\lambda^c$ based on the optimized value of $x$ and parameterized value of $d$ and electricity tariff $\lambda^u$. Finally, the insurer returns the optimal value of $x$ considering the optimized value of $\lambda^c$ as a constraint. Notably, in the TL optimization in Fig.~\ref{fig:tri-level problem}, $\lambda^u$ is optimized in the LL. This implies that the premium based on the TL optimization is derived using the optimized values of the $\lambda^c$ and $\lambda^u$. Next, using Eqs.~\eqref{eq:objective_upper_bi}-\eqref{eq:EVCS_cost}, we state and prove:

\begin{theorem}
\label{the:theorem1}
\textup{ Consider the model in Eqs.~\eqref{eq:ul}-\eqref{eq:ml}  and let $\lambda_t^{u,*},~\forall t \in \mathcal{T},$ be a predetermined electricity tariff for the insurance period (e.g., \cite{con_ED}). Then, the optimal  cyber insurance premium follows as:}
\begin{align}
      x^{*} \!\!=&\frac{B}{[(\mathbb{P}(A)(\!\gamma\!-\!1\!)\! +\!1\!)\! -\!B]}\!\! \bigg[\!\mathbb{P}(A)\rho\!\!\sum_t\!\! D_t  
     \!\!+\!\! (1\!-\!\mathbb{P}(A)) \!\!\sum \!\!D_t \lambda_t^{u,\!*} \!\bigg], 
    \label{eq:x_analytical}
\end{align}
\textup{where parameters  $ B= \frac{\mathbb{P}(A)\gamma (1+\kappa A_{h})}{1-r}$ and $D_t=\sum_s \phi^s d_t^s$.}

\textit{Proof:}
\textup{As per Eq.~\eqref{eq:EVCS_cost}, the EVCS  breaks even when $\lambda_t^c$  is such that Eq.~\eqref{eq:EVCS_cost} is binding. Thus, the Lagrangian function of the LL problem is:}
\begin{align}
\notag
    \mathcal{L} =&\sum_t (\lambda_t^{c})^2  + \omega\bigg[(1- \mathbb{P}(A))\sum_{t=1}^{T}\bigg(D_t(\lambda_t^{u,*} -\lambda_t^{c})\bigg)  \label{eq:lag_EVCS}  \\ 
    &+\sum_{t}^{T} (\hat x^{*}D_{t} )  + \mathbb{P}(A) \sum_{t=1}^{T}\bigg( D_{t}( \rho -\gamma \lambda_t^{c} \bigg) \bigg],
\end{align}
\textup{where $\omega$ is a Lagrangian dual variable associated with the binding constraint in Eq.~\eqref{eq:EVCS_cost}. Next, the first-order optimally conditions with respect to $\lambda_t^c,  \forall t \in \mathcal{T},$ is:}
\begin{align}
     \frac{\partial{L}}{\partial{\lambda^c_t}}&= 2\lambda_t^c - \omega \bigg(\mathbb{P}(A)\gamma D_t +(1-\mathbb{P}(A)) D_t\bigg)=0. \label{eq:lag_lower_dt} 
\end{align}
\textup{Solving Eq.~\eqref{eq:lag_lower_dt} for $\lambda^{c}_t$ leads to:}
\begin{align}
\label{eq:lambda_c_opt}
    &\lambda^{c,*}_t = \frac{1}{2}\omega\bigg[(\mathbb{P}(A)(\gamma-1)+1)D_t \bigg], \quad \forall t \in 1:T.
\end{align}
\textup{In turn, $\omega$ can be expressed from setting $\frac{\partial{L}}{\partial{w}} =0$ and comparing the result with Eq.~\eqref{eq:lambda_c_opt}, which returns:}

\begin{align}
    \omega = &\frac{2}{\sum_t D_t^2[\mathbb{P}(A)(\gamma -1)+1]^2}\bigg[\bigg(\mathbb{P}(A)\rho + \lambda_t^{u,*}(1-\mathbb{P}(A)) \notag \\ \label{eq:w}
    &+\hat x^{*}\bigg) \sum_t D_t \bigg].
\end{align}

\textup{Next, the expression for $\lambda_t^{c,*}$ in Eq.~\eqref{eq:lambda_c_opt} is used to solve the UL optimization in Eq.~\eqref{eq:ul}.
When solving the optimization in Eq.~\eqref{eq:ul},  Eq.~\eqref{eq:premium_bound} becomes binding to  attain the least-cost objective value and, thus, the Lagrangian function of the LL problem in  Eq.~\eqref{eq:ul} follows as:}
\begin{align}
\label{eq:lag_upper}
   \mathcal{L} &=x + \delta \bigg[B\sum_{t} D_{t}\lambda_{t}^{c,*} -x\bigg],
\end{align}
\textup{where $\delta$ is a Lagrangian  dual variable of Eq.~\eqref{eq:premium_bound}. Using  $\lambda_t^{c*}$ in Eq.~\eqref{eq:lambda_c_opt} and then Eq.~\eqref{eq:lambda_c_opt}  in  Eq.~\eqref{eq:lag_upper}, the first-order optimality condition  with respect to $\delta$ is:}

\begin{align}
  \label{eq:lag_upper_w}  
  \frac{\partial{L}}{\partial{\delta}} =& \frac{1}{2}\omega B \sum_{t}\bigg[D_{t} \bigg((\mathbb{P}(A)(\gamma-1)+1) D_t\bigg)\bigg] -x=0.
\end{align}
\textup{Solving Eq.~\eqref{eq:lag_upper_w} for $x^{*}$ gives Eq.~\eqref{eq:x_analytical}}. \qed 
\end{theorem}

\begin{rem}
The cyber insurance premium in Eq.~\eqref{eq:x_analytical}, which is obtained from Eqs.~\eqref{eq:ul}-\eqref{eq:ml}, is directly proportional to the  expected EVCS demand forecast, i.e., $\sum_s \phi^s d_t^s = D_t$. Thus, the EVCS operator can reduce its premium by strategically reporting a lower value of the EVCS demand forecast. However, we assume this premium reduction is very unlikely because the commercial EVCSs report their demand forecast to multiple entities (e.g., power grid operator, insurer, public users), and hence, the insurer may cross-validate the reported forecasts.
\end{rem}

\subsection{DLMP-based Tariff}
\label{sec:tri}
As opposed to the predetermined electricity tariff underlying the analytical solution in Theorem 1, electric power distribution systems are expected to gradually transition towards dynamic and more spatially and temporally granular tariffs (e.g., by means of the DLMPs \cite{caramanis2016co}). To account for this eventuality, the optimization in   Eqs.~\eqref{eq:objective_upper_bi}-\eqref{eq:EVCS_cost} is recast as the TL optimization model  shown in Fig.~\ref{fig:tri-level problem}. In this TL optimization,  the UL and Middle-Level (ML) problems are the insurer and EVCS with the roles explained above, while the LL is  a model of the distribution system operator used to compute DLMPs. Mathematically, the UL and ML optimizations are given by   Eq.~\eqref{eq:UL-tri}  and Eq.~\eqref{eq:ML-tri}, respectively, while the LL optimization is given in   Eqs.~\eqref{eq:objective_lower_tri}-\eqref{eq:load_const}. The resulting TL optimization   follows then as:
\begin{subequations}
\begin{align}
    \label{eq:UL-tri}
    & \text{Eqs.~ \eqref{eq:objective_upper_bi}-\eqref{eq:premium_bound}} \hspace{13mm}: \text{UL problem} \\
    \label{eq:ML-tri}
     & \text{Eqs.~ \eqref{eq:objective_lower_bi}-\eqref{eq:EVCS_cost}}  \hspace{13mm}: \text{ML problem} \\
   \label{eq:objective_lower_tri}
   &\boldsymbol{\lambda^{u,*}} = \argmin_{\Xi^{G}}  ~ C^s_{LL}:= \sum_{t}\sum_{i}C_i^g  g_{i,t}^{s}, ~ \forall s \in \mathcal{S} \\ \label{eq:gen_const} 
   & \quad s.t.~ 
   0\leq g_{i,t}^{s} \leq \overline{G}_i :(\underline{\alpha}_{i,t}^{s} ,\overline{\alpha}_{i,t}^{s}), ~ \forall s \in \mathcal{S}, ~\forall t \in \mathcal{T}, ~\forall i \in \mathcal{I},  \\
   \label{eq:flow_const}
   &\quad f_{l,t}^{s} z_{l} =\theta_{o(l),t}^{s}-\theta_{r(l),t}^{s} 
   : (\xi_{l,t}^{s}), ~ \forall s \in \mathcal{S}, ~\forall t \in \mathcal{T}, ~\forall l \in \mathcal{L}, \\ 
  \label{eq:line_const}  
   &\quad -\overline{F}_l \leq f_{l,t}^{s} \leq \overline{F}_l : (\underline{\delta}_{l,t}^{s},\overline{\delta}_{l,t}^{s}), \forall s \in \mathcal{S}, ~\forall t \in \mathcal{T},~\forall l \in \mathcal{L}, \\
   &\quad d_{b,t}^{d,s} = \sum_{i \in I_b}g_{i,t}^{s} +\sum_{l|r(l)=b}f_{l,t}^{s} -\sum_{l|o(l)=b}f_{l,t}^{s} -d_{b,t}^{s}: (\lambda_{b,t}^{s}), \notag \\ \label{eq:load_const}
   & \hspace{14mm} ~\forall s \in \mathcal{S},~\forall b \in \mathcal{B}, ~ \forall t \in \mathcal{T},
\end{align}
\end{subequations}
 where ${\Xi^{G}} \!\!= \!\!\{g_{i,t}^{s}, ~\theta_{n,t}^{s}, ~f_{l,t}^{s} \in \mathbb{R}\!\!:\! g_{i,t}^{s}, ~\theta_{n,t}^{s}, ~f_{l,t}^{s}\geq 0\}$  and $\boldsymbol {\lambda^{u}} = [\lambda_{b,t}^s, \ldots, \lambda_{b,T}^s]^{\top}$.
 The objective of the LL optimization in  Eq.~\eqref{eq:objective_lower_tri} is to minimize the expected generation cost $(C_{LL})$ in the electric power distribution system. Eq.~\eqref{eq:gen_const} enforces power generation limits on the generators, Eq.~\eqref{eq:flow_const} models DC power flows, Eq.~\eqref{eq:line_const} enforces  flow capacity limits, and Eq.~\eqref{eq:load_const} ensures the nodal power balance. The  dual variable of Eq.~\eqref{eq:load_const}, $\lambda_{b,t}^s$, is adopted as a DLMP  at bus $b$ for time $t$ and day $s$. We model loads  as  price-takers, assume the  total  system  load is given, and  use  DC power  flows  as  a  proxy  to  compute  DLMPs,  but  other  power flow  formulations  (e.g.,  LinDistFlow \cite{baran1989optimal}) and flexible loads can be adopted to compute additional DLMP components.

The TL optimization in Eq.~\eqref{eq:UL-tri}-\eqref{eq:load_const} does not have an analytical solution and can only be solved numerically, e.g., by means of  the C\&CG algorithm, \cite{zeng2013solving, ruiz2015robust}, described  in Section~\ref{sec:CCG_tri_robust}. However, the use of this algorithm requires a reformulation of the TL optimization as an equivalent BL problem.  Since the LL problem is convex and linear, the  strong duality condition holds and  the single-level equivalent of the ML and LL can be obtained using the Dual  LL (DLL) formulation given  as ($\forall s \in \mathcal{S}$):
\begin{subequations}
\begin{align}
\notag
\label{eq:objective_dll}
    &\min_{\Xi^{DLL}} C_{DLL}^{s}:= \sum_{t \in \mathcal{T}}\bigg(\sum_{i \in \mathcal{I}}\overline{\alpha}_{i,t}^{s}\overline{G}_i +\sum_{i \in \mathcal{L}} (\overline{\delta}_{l,t}^{s}+\underline{\delta}_{l,t}^{s})\overline{F}_l \\ 
    & \hspace{20mm} - \sum_{b \in \mathcal{B}}(d_{b,t}^{d,s} + d_{b,t}^{s})\lambda_{b,t}^{s} \bigg) \\ 
    \label{eq:dll_constraint1}
    & s.t. \quad  \overline{\alpha}_{i,t}^{s} -\underline{\alpha}_{i,t}^{s} -\lambda_{b(i),t}^{s} =-C_i^g, \quad \forall i \in \mathcal{I},~ \forall t \in \mathcal{T}, \\ \label{eq:dll_constraint2}
    & \xi_{l,t}^{s} + \overline{\delta}_{l,t}^{s} - \underline{\delta}_{l,t}^{s}+\lambda_{o(l),t}^{s}-\lambda_{r(l),t}^{s} =0,~ \forall l \in \mathcal{L},~ \forall t \in \mathcal{T}, \\ \label{eq:dll_constraint3}
    & \sum_{l\mid r(l)=b} \frac{\xi_{l,t}^{s}}{z_l} -\sum_{l\mid o(l)=b} \frac{\xi_{l,t}^{s}}{z_l} =0 , \quad \forall l \in \mathcal{L}, ~\forall t \in \mathcal{T}, 
\end{align}
\end{subequations}
where ${\Xi}^{DLL} = \{\overline{\alpha}_{i,t}^{s}, \underline{\alpha}_{i,t}^{s}, \overline{\delta}_{l,t}^{s},\underline{\delta}_{l,t}^{s} \geq 0, \xi_{l,t}^{s}, \lambda_{b,t}^{s} :\text{free} \}$. Given Eq.~\eqref{eq:objective_dll}-\eqref{eq:dll_constraint3}, the TL model is equivalent to:
\begin{subequations}
\begin{align}
\label{eq:tl_equivalent_ul}
    &\text{Eq.~\eqref{eq:objective_upper_bi}}\hspace{23.5mm} : \text{UL problem} \\ \notag
    &\text{s.t~ Eqs.~\eqref{eq:loss_revenue}-\eqref{eq:premium_bound}}, \\
    &\lambda_{t}^c  \in \argmin\{ \\
    &\text{Eqs.~\eqref{eq:objective_lower_bi}-\eqref{eq:EVCS_cost}}, \hspace{14.5mm} : \text{ML problem} \\
    & \text{Eqs.~\eqref{eq:gen_const}-\eqref{eq:load_const},} \hspace{10.5mm}: \text{LL primal constraints} \\
    &\text{Eqs.~\eqref{eq:dll_constraint1}-\eqref{eq:dll_constraint3}}, \hspace{11mm} : \text{LL dual constraints} \\ \label{eq:tl_equivalent_sd}
    & O_{LL}^{s} = O_{DLL}^{s}, ~\forall s \in \mathcal{S} \hspace{2mm} : \text{LL strong duality}\}.
\end{align}
\end{subequations}
The  BL equivalent in Eqs.~\eqref{eq:tl_equivalent_ul}-\eqref{eq:tl_equivalent_sd} is then solved using the C\&CG algorithm as explained in  Section~\ref{sec:CCG_tri_robust}.

\section{Estimating \texorpdfstring{$\mathbb{P}(A)$}{P(A)} Using a Semi-Markov Process}
\label{sec:SMP}

\begin{figure}[!t]
  \vspace{-2mm}
\begin{tikzpicture}[font=\sffamily]
        \node[state, 
            minimum width=0.5cm,
            draw=none,
            line width=0.5mm,
            fill=gray!50!green] at (0, 0)     (Good)     {G};
          
        \node[state, 
            minimum width=0.5cm,
            draw=none,
            line width=0.5mm,
            fill=gray!80!red] at (3, 0)    (Intrusion)     {I};
        \node[state, 
            minimum width=0.5cm,
            draw=none,
            line width=0.5mm,
            fill=gray!90!green] at (6, 0)     (Detection)     {D};
        \node[state, 
            minimum width=0.5cm,
            draw=none,
            line width=0.5mm,
            fill=gray!50!red] at (7.7, 0)     (Failure)     {F};
        \node[state, 
            minimum width=0.5cm,
            draw=none,
            line width=0.5mm,
            fill=gray!95!green] at (3, -1.2) (Containment) {C};
        \draw[every loop,
              auto=right,
              line width=0.2mm,
              >=latex,
              draw=black,
              fill=black]
            (Good)     edge[bend left=60]            node {$H_{GI}$} (Intrusion)
            (Intrusion)     edge[bend left=30] node {$H_{ID}$} (Detection)
            (Intrusion)     edge[bend left=50]            node {$H_{IF}$} (Failure)
            (Detection)     edge[bend left=30] node {$H_{DC}$} (Containment)
            (Containment) edge[bend left=20]            node {$H_{CG}$} (Good)
            (Failure) edge[bend left=55, auto=right] node {$H_{FG}$} (Good);
    \end{tikzpicture}
    \vspace{-3mm}
  \caption{Semi-Markov Process (SMP) for cyberattacks on EVCSs. $H_{(\cdot)}$ defines the Cumulative Distribution Function (CDF) of the state transition time.}
  \label{fig:mc}
\end{figure}
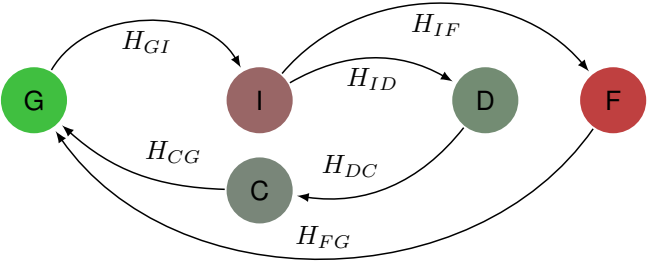
The optimization problems in Eqs.~\eqref{eq:objective_upper_bi}-\eqref{eq:EVCS_cost} and  Eqs.~\eqref{eq:tl_equivalent_ul}-\eqref{eq:tl_equivalent_sd} require estimation of insurance policy factors $\gamma$, $\kappa$, $r$, and $\mathbb{P}(A)$. While $\gamma$, $\kappa$, and $r$ can be chosen based on the current insurance industry practice (see \cite{fitch,romanosky2019content}),  $\mathbb{P}(A)$ must be estimated. We use Markovian state transitions shown in Fig.~\ref{fig:mc} to estimate $\mathbb{P}(A)$. 
We consider five states that represent  cyberattacks on the EVCS and recovery \cite{liu2019probabilistic, lau2020cybersecurity}: Good (G), Intrusion (I), Detection (D), Containment (C), and Failure (F).  During normal operations, the EVCS is in state G. State I indicates that the EVCS and/or its server are infected malicious actions (see the attack routes in \cite{acharya2020cybersecurity} and threat analysis such as \cite{lee2015electric}). After the infection, the attack is either detected in state D, which returns the EVCS to state G via state C, or remains undetected, which causes failure of the EVCS operation given by state F. Once the EVCS detects the attack, it may contain the attack to avoid it from spreading  at state C before returning the EVCS to state G.  Following the success of the attack at state F, the  EVCS recovers  to state G. 

\begin{rem}
Although states G, I, D, C and F are generic for any cyber-physical system, the data-driven state transitions are specific to EVCSs. For example, state I includes all malicious actions in EVCSs, e.g., phishing emails, SQL injections and password intrusion. Use of EVCS attack data in the SMP given in Fig.~\ref{fig:mc} makes it specific to cyberattacks on EVCSs. 
\end{rem}

Let $\mathcal{S}=\{G, I, D, C, F \}$ be a state space  of the Markov chain representing the effect of cyberattacks on the EVCS and let  $X_n \in \mathcal{S}, ~ n \in \mathcal{N} =\{0,1,2,\ldots\}$ denote the states of the cyberattack after $n$ transitions in the Markov chain. The state transitions occur at jump times $T_n$. Thus, the SMP is a sequence of two-dimensional random variables $(X_n, T_n)$, where the  attack states $G, I, D, C, \text{and}~F$  are stochastic over time. The SMP evolves over continuous time $t \in [T_n, T_{n+1})$ and is formalized  as:
\begin{align}
    \notag
    &\mathbb{P}(X_{n+1} =j, \tau_{n+1}\leq t \mid (X_0,T_0),\ldots, (X_n=i, T_n))\\ 
    &=\mathbb{P}(X_{n+1} =j, \tau_{n+1}\leq t \mid X_n =i) \\ \notag
    &=\mathbb{P}_{ij}H_{ij}(t), \quad \forall n \in \mathcal{N}, ~\forall i,j \in \mathcal{S},
\end{align}
where $\tau_{n+1} = T_{n+1}-T_n$ is the sojourn time, i.e., the time spent in a given state between transitions $n$ and $n+1$,  $\mathbb{P}_{ij}$ is the probability of transitioning from state $i$ to $j$,  
and $H_{ij}(t)$ is the Cumulative Distribution Function (CDF) of the transition time between   states $i$ and $j$. Below we provide a two-stage method to solve the SMP based on \cite{kumar2013availability, liu2019probabilistic}, where the first stage returns the transition probability of the embedded Markov chain of the SMP and the second stage returns the steady-state probabilities of the SMP states. The embedded Markov chain ignores the time spent at a state and only accounts for the transition probability between the states. 

\subsubsection{Stage 1}
The kernel matrix denoted as $K(t)$ is given as:
\begin{align}
\label{eq:kernel}
   K(t) = \begin{bmatrix}
    0 & k_{GI} & 0 & 0 &0\\
    0 & 0 & k_{ID}& 0& k_{IF}\\
    0 & 0 & 0 & k_{DC}& 0\\
    k_{CG} & 0 & 0 & 0 & 0\\
    k_{FG} & 0 & 0 & 0 & 0 \\
    \end{bmatrix}.
\end{align}
Assuming a Weibull distribution of state transitions given as $H_{ij}(t, \alpha_{ij}, \beta_{ij}) = 1-\exp{-(\frac{t}{\alpha_{ij}})^{\beta_{ij}}}$, the elements of $K(t)$ in Eq.~\eqref{eq:kernel} are determined as:
\begin{equation}
\begin{split}
      & k_{GI}(t) =  H_{GI}(t), \quad  k_{ID}(t) = \int_{0}^{t}\bar{H}_{IF}(t)dH_{ID}(t) \\
    & k_{IF}(t) = \int_{0}^{t}\bar{H}_{ID}(t)dH_{IF}(t), \quad k_{DC}(t) = H_{DC}(t),\\
    & k_{CG} (t) = H_{CG}(t), \quad  k_{FG} (t) = H_{FG}(t),  
\end{split}
\label{eq:kgg}
\end{equation}
where $\bar{H}_{ij} =1-H_{ij}$. The Weibull distribution parameters of  state transitions (i.e., $\alpha, \beta$)  are obtained by fitting empirical data from historical cyberattacks.

\begin{rem}
Due to limited publicly available data on EVCS cyberattacks, we model the Weibull distribution parameters using data from cyberattacks in the information technology industry, which is consistent with the current real-world practice. The data on cyberattacks that we use is explained in Section~\ref{sec:case_study}. The studies in  \cite{schroeder2009large, woods2019county, franke2014distribution}
support the use of Weibull, Lognormal, Gamma, and Exponential distributions for modeling the cyberattacks. Among these distributions, we chose a Weibull distribution because of its best fit on our data.
\end{rem}

The steady-state probability matrix of the embedded Markov chain 
is determined for $t \to \infty$.  Thus,  we state the kernel matrix in Eq.~\eqref{eq:kernel} as $K(t \to \infty) =1$, which means that the probability of transitioning between states  approaches 1 in infinite time.  Since the embedded Markov chain  is  discrete-time, applying the steady-state condition to the kernel matrix, leads to the vector of steady-state probabilities of the embedded Markov chain given as $\boldsymbol{p} = [ p_G~ p_I~ p_D~ p_C~  p_F]$:

\begin{equation}
\label{eq:p}
    \boldsymbol{p}=\boldsymbol{p} K^{\top}(t \to \infty); \quad \boldsymbol{p}\boldsymbol{e}^\top =1,
\end{equation}
where $\boldsymbol{e}= [ 1~ 1~ 1~ 1~1].$

\subsubsection{Stage 2}
This stage calculates  sojourn time $T_{s}$  spent in  state $s$ before transitioning to another state: 

\begin{equation}
\begin{split}
      &T_{G} = \int_{0}^{\infty}\bar{H}_{GI}(t) dt, \quad T_{I} = \int_{0}^{\infty}\bar{H}_{ID}(t) \bar{H}_{IF}dt, \\
    &T_{D} = \int_{0}^{\infty}\bar{H}_{DC}(t)dt, \quad T_{C} = \int_{0}^{\infty}\bar{H}_{CG}(t)dt,  \\
    &T_{F} = \int_{0}^{\infty}\bar{H}_{FG}(t)dt.   
\end{split}
\label{eq:T}
\end{equation}

Let $\boldsymbol{T}= [ T_G~T_I~ T_D~T_C~T_F]^{\top}$. Using the values of $\boldsymbol{p}$ in Eq.~\eqref{eq:p} and $\boldsymbol{T}$ in Eq.~\eqref{eq:T}, the steady-state probability of transitioning to  failure state $F$, e.g.,  the probability of  attack success  $\mathbb{P}(A)$, is: 
\begin{equation}
    \mathbb{P}(A) = \mathbb{P}_{F} = \frac{p_{F}T_{F}}{\boldsymbol{p}\boldsymbol{T}^{\top}}.
\end{equation}
 
Since the estimation of $\mathbb{P}(A)$ invokes dealing with uncertainty, Section~\ref{sec:robust} robustifies the optimizations in  Section~\ref{sec:modesl} against the uncertainty in $\mathbb{P}(A)$.

\section{Distributionally Robust and Risk- Averse Insurance Premium Design}
\label{sec:robust}
This section extends the optimizations in Section~\ref{sec:modesl} to account for  the uncertainty of insurance policy factors $\Pi$ and risk-averse EVCS attitudes. 

\subsection{Robust and Risk-Averse Bi-Level Optimization Model}
Inspired by the analytical expression in  Eq.~\eqref{eq:x_analytical} of Theorem~\ref{the:theorem1}, we seek to robustify  this result against uncertain policy factors $\mathbb{P}(A)$, $r$, and $\kappa$. To this end,  we first assume that any parameter estimation error is contained in a box-constrained confidence interval as practiced by the industry \cite{romanosky2019content, fitch}, where the box limits can be elicited using expert knowledge, industry practices, and/or statistical analysis.
Second, we revisit Eqs.~\eqref{eq:objective_lower_bi}-\eqref{eq:EVCS_cost} to include a CVaR  metric as a means of considering EVCS risk attitudes.

\subsubsection{Robust and Risk-Averse EVCS Optimization Problem}
\label{sec:EVCS_robust}
Parameter $\mathbb{P}(A)$ in the EVCS optimization  in Eq.~\eqref{eq:ml} is subject to an  error caused by potential data insufficiency (e.g., limited  samples could be available). We model this error as  random variable $\epsilon_p$ such that $\mathbb{P}(A) := \overline{\mathbb{P}}(A) \pm \epsilon_p$, where  $\overline{\mathbb{P}}(A)$ is an estimated value of the steady-state attack probability in Section~\ref{sec:SMP}.  Random variable $\epsilon_p$  captures such errors (e.g., modeling errors and erroneous data) and allows for the insurers to internalize their confidence in the calculations of $\mathbb{P}(A)$. 
Note that  the subscript p in $(\cdot)_p$ refers to the parameter $\mathbb{P}(A)$. Accordingly,  the  box-constrained confidence interval is given as $\mathbb{P}(A)\in [\underline \Gamma_p, ~ \overline \Gamma_p ]$, where $\overline \Gamma_p = \overline{\mathbb{P}}(A) + \epsilon_p$  and $\underline \Gamma_p = \overline{\mathbb{P}}(A) - \epsilon_p$.
Assume $\epsilon_p$ follows a normal distribution with zero mean $\mathbb{E}({\epsilon_p)}=0$ and variance $\text{Var}(\epsilon_p)=\sigma_p^2$. 
Then, the box constraint can be defined as :
\begin{equation} 
    \label{eq:UL_dro_bound1}
    \underline{\Gamma}_p = \overline{\mathbb{P}}(A) -t_{(1-\xi/2)}\frac{\sigma_p}{\sqrt{N}},~ \overline{\Gamma}_p = \overline{\mathbb{P}}(A) +t_{(1-\xi/2)}\frac{\sigma_p}{\sqrt{N}},
   \end{equation}
where $\overline{\mathbb{P}}(A)$ is an empirical mean modeled by the $t$-distribution, $t_{(1-\xi/2)}$ is the ${(1-\xi/2)}$-quantile of the $t$-distribution computed from  $N$ observations of ${\mathbb{P}}(A)$.

Using Eq.~\eqref{eq:UL_dro_bound1}, the risk-averse formulation of the EVCS optimization problem in Eq.~\eqref{eq:ml} can be recast as follows:
\begin{subequations}
\label{eq:opt_EVCS_CVaR}
\begin{align}
\label{eq:CVAR_EVCS_obj}
    \min_{{\Xi^{E}}} &~||{\boldsymbol{\lambda^c}}||_2^2 \\ \label{eq:CVAR_EVCS_const}
         & s.t.~ \text{CVaR}_\alpha(C(d^s, \phi^s, \lambda^c)) \leq 0,
\end{align}
\end{subequations} 
where $\text{CVaR}_{\alpha}$ of the EVCS cost $C(d^s,\phi^s, \lambda^c)$, with  probability distribution $\phi^s$, is defined as the mean of $\alpha-$tail distribution of  $C(d^s,\phi^s, \lambda^c)$ \cite{rockafellar2007coherent, hans2019risk}, that is:
\begin{subequations}
\label{eq:EVCS_CVaR}
\begin{align}
\label{eq:EVCS_CVaR_obj}
    &\text{CVaR}_\alpha(C(d^s, \phi^s, \lambda^c)) = \sup_{\tilde \phi \in Q_\alpha}\sum_{s \in S}\tilde \phi^s\tilde C(d^s, \lambda^c),
\end{align}
where:
\begin{align}
    &Q_\alpha = \{\tilde \phi:\sum_s \tilde\phi^s =1, ~ \tilde\phi^s \le \frac{1}{\alpha} \phi^s \}, \\
    \tilde C(d^s, \lambda^c) = & (1- \mathbb{P}(A))\sum_{t=1}^{T}d_{t}^{s}(\lambda_t^{u,*} -\lambda_t^{c}) + \sum_{t=1}^{T} \hat x^{*} d_t^s \notag \\ \label{eq:cost_CVAR} 
    & +\mathbb{P}(A) \sum_{t=1}^{T}(d_{t}^{s} (\rho-\gamma \lambda_t^{c})) .
\end{align}
Since $\mathbb{P}(A)$ holds a linear relationship with the EVCS cost in Eq.~\eqref{eq:cost_CVAR}, the upper bound of $\mathbb{P}(A)$ in Eq.~\eqref{eq:UL_dro_bound1} gives the \mbox{$\alpha$-worst-case} (\mbox{$\alpha$-WC}) estimation of the  EVCS cost. Thus, Eq.~\eqref{eq:cost_CVAR} is recast as:
\begin{align}
    \tilde C^{WC}(d^s, \lambda^c) = & (1- \overline \Gamma_p)\sum_{t=1}^{T}d_{t}^{s}(\lambda_t^{u*} -\lambda_t^{c}) + \sum_{t=1}^{T} \hat x^{*} d_t^s \notag \\ \label{eq:cost_CVAR_WC} 
    & +\overline \Gamma_p \sum_{t=1}^{T}(d_{t}^{s} (\rho-\gamma \lambda_t^{c})). 
\end{align}
\end{subequations}
Using Eq.~\eqref{eq:cost_CVAR_WC} in the EVCS optimization in  Eq.~\eqref{eq:opt_EVCS_CVaR}, the EVCS operator designs a minimum $\lambda_t^{c, WC}, ~ \forall t \in \mathcal{T}$, that at least recovers the \mbox{$\alpha$-WC} EVCS cost.
In this setting, the risk-averse EVCS optimization problem in Eq.~\eqref{eq:opt_EVCS_CVaR} follows:
\begin{subequations}
\label{eq:EVCS_CVAR_primal}
\begin{align}
\label{eq:EVCS_CVAR_primal_obj}
    \min_{\lambda^{c,WC}, \zeta, v} & ~||{\boldsymbol{\lambda^{c,WC}}}||_2^2  \\
    \label{eq:primal_cvar_constraint1}
         & s.t.~ v + \sum_s \phi^s \zeta_s \leq 0,: (\eta) \\
         &\alpha \zeta^s \geq \tilde C^{WC}(d^s, \lambda^{c}) -v,: (\varphi^s), ~ \forall s \in 1:S, \\
         & \zeta^s \geq 0, : (\mu^s)  ~ \forall s \in 1:S, \label{eq:primal_cvar_constraint} \\ \label{eq:primal_cvar_constraint_last}
         & \lambda_t^{c,WC} \geq 0, : (\beta_t) ~ \forall t \in 1:T,
\end{align}
\end{subequations}
where $\boldsymbol{\lambda^{c,WC}} = [\lambda_{1}^{c,WC} \ldots, \lambda_T^{c,WC}]^{\top} $, $v$ is an auxiliary variable used to recast Eq.~\eqref{eq:EVCS_CVaR_obj} as an minimization problem and $\eta$, $\varphi^s$, $\beta_t$, and $\mu^s$  are the Lagrangian dual variables of  Eq.~(\ref{eq:EVCS_CVAR_primal}b)-(\ref{eq:EVCS_CVAR_primal}e). Next, using the KKT conditions of the optimization in Eq.~\eqref{eq:EVCS_CVAR_primal}:   
\begin{subequations}
\label{eq:KKT}
\begin{align}
   & v + \sum_s \phi^s \zeta^s \leq 0 \\
   & -\alpha \zeta^s -v + \tilde C^{WC}(d^s,\lambda^c) \leq 0, \quad s \in 1:S, \\
   & -\zeta^s \leq 0, ~\varphi^s \geq 0, \quad s \in 1:S,\\
   & \eta \geq 0, ~\mu \geq 0, \beta_t \geq 0,\quad t \in 1:T, \\
   & \eta(v+ \sum_s \phi^s \zeta^s) = 0, \\
   & \varphi^s(-\alpha\zeta^s -v + \tilde C^{WC}(d^s,\lambda^c)) =0, \quad s \in 1:S, \\
   & \mu ^s (-\zeta^s) =0 , \quad s \in 1:S, \label{eq:zeta_positivity_cond}\\
   & \eta\phi^s -\alpha \varphi^s - \mu^s =0 , \quad s \in 1:S, \label{eq:optcond_zeta_s}\\
   & \beta_t(-\lambda_t^{c,WC}) = 0, \quad t \in 1:T,\\
   \label{eq:kkt_eta}
   & \eta -\sum_s \varphi^s =0, \\
   \label{eq:lambda_WC}
   & 2\lambda_t^{c,WC} - [\overline \Gamma_p (\gamma -1) +1]\sum_s (\varphi^s d_t^s) -\beta_t = 0 , \quad t \in 1:T.
\end{align}
\end{subequations}
we finally obtain:
\begin{align}
    (1-\alpha)\sum_s\varphi^s = \sum_s\mu^s. \label{eq:cvar_dual}
\end{align}
This relationship captures the trade-off solved by the CVaR between adding scenarios to the set of worst-cases and adjusting their relative probability. 
For example, if $\alpha=1$, then Eq.~\eqref{eq:cvar_dual} yields $\sum_s\mu^s = 0$, which implies $\mu^s = 0, \forall s \in \mathcal S,$ and $\zeta^s\ge0, \forall s \in \mathcal S$ as per Eq.~\eqref{eq:zeta_positivity_cond}. In other words, for $\alpha=1$ \textit{all} scenarios are considered with their original probability $\phi^s$ such that $\sum_s \phi^s \zeta_s$ recovers EVCS cost \textit{in expectation} as in Section~\ref{sec:bilevel_equiv_insurance_premium}.
On the other hand, if $\alpha=0$, Eq.~\eqref{eq:EVCS_CVAR_primal} only considers the single WC scenario as $\zeta_s=0$ and $v \geq \tilde C^{WC}(d^s, \lambda^c), \forall s \in \mathcal{S}$, as per Eq.~\eqref{eq:primal_cvar_constraint}. Since $\sum_s \varphi_s = \eta$ in Eq.~\eqref{eq:kkt_eta}, $\varphi^s$ is a proxy for the probability of typical days. Thus, we define a relative probability of typical day $s$ as $\hat \varphi_s =\frac{\varphi_s}{\eta}, ~ \forall s \in S$. Given $\hat \varphi_s$, Eq.~\eqref{eq:EVCS_CVAR_primal} determines ${\lambda_t}^{c,WC}, \forall t \in \mathcal{T},$ based on the \mbox{$\alpha$-WC} EVCS cost.

\subsubsection{Robust Insurer Optimization Problem}
\label{sec:insurer_robust}
 Given the value of  $\lambda_t^{c, WC}$ and EVCS demand profiles $d_t^s~ \forall s \in \mathcal{S}, ~ \forall t \in \mathcal{T}$,
 we robustify the insurer optimization  by introducing box constraints on uncertain  parameters  $\mathbb{P}(A)$, $\kappa$, and $r$. The insurer uses the box constraint for $\mathbb{P}(A)$ as described in Eq.~\eqref{eq:UL_dro_bound1}.  Consider the box constraints for $\kappa$ and $r$ as $[\underline \kappa, ~ \overline \kappa]$ and $[\underline r, ~ \overline r]$. Since $\kappa$ and $r$ are industry-specific  parameters, the range for the box constraint is industry-specific \cite{romanosky2019content}. Accordingly, the insurer optimization problem  in Eq.~\eqref{eq:ul} is recast as:
\begin{subequations}
\label{eq:ul_robust} 
\begin{align} \label{eq:objective_upper_bi_robust} 
    \min &~ x^{WC}   \\ \label{eq:loss_revenue_robust}
    s.t.&~ CL = \underbrace{\frac{\gamma (1 + \overline \kappa A_h)\overline \Gamma_p}{1- \overline r}}_{\text{policy factors}}\underbrace{\bigg(\sum_{s=1}^{S} \phi^{s}\sum_{t=1} ^{T}d_{t}^{s}\lambda_t^{c, WC}\bigg)}_{\text{EVCS revenue}} \\  \label{eq:premium_bound_robust}
    & x \geq CL.
    \end{align}
\end{subequations}
The risk-averse EVCS has $\hat \varphi^s$ as the relative probability of typical days. However, the insurer in Eq.~\eqref{eq:ul_robust} may not have the same information on $\hat \varphi^s$ as the EVCS (see Remark 1 in Section~\ref{sec:modesl}). For example, the insurer optimizes the premium using the expected EV charging demand profiles during the insured period, but the EVCS optimization uses  WC typical days instead.
Therefore, an exact analytical expression for the insurance premium cannot be obtained in this case, but it can be determined numerically. Since the risk-averse LL problem in Eq.~\eqref{eq:EVCS_CVAR_primal} is convex, we  convert the robust and risk-averse BL optimizations, insurer in Eq.~\eqref{eq:ul_robust} and EVCS in Eq.~\eqref{eq:EVCS_CVAR_primal}, to the following single-level equivalent \cite{gumucs2001global}:
\begin{subequations}
\label{eq:bi-level-CVaR}
\begin{align}
    & \text{Eq.~}\eqref{eq:ul_robust}, \quad \text{: UL problem} \\
    &  \text{Eq.~}\eqref{eq:EVCS_CVAR_primal}, \quad \text{: LL primal problem} \\
    &  \text{Eq.~}\eqref{eq:KKT}, \quad \text{: KKT conditions for ML Problem} 
\end{align}
\end{subequations}
The insurance premium for \mbox{$\alpha$-WC} EVCS cost is thus obtained by numerically solving the single-level equivalent in Eq.~\eqref{eq:bi-level-CVaR}.

\subsection{Robust and Risk-Averse Tri-Level Optimization Model}
\label{sec:CCG_tri_robust}
This section recasts the original TL optimization in case of the robust  insurer and EVCS presented in Sections~\ref{sec:EVCS_robust} and \ref{sec:insurer_robust}. Thus, the TL optimization in Section~\ref{sec:tri} results in: 
\begin{subequations}
\label{eq:tri_robust}
\begin{align}
    &\text{Eq.~}\eqref{eq:objective_upper_bi_robust} \hspace{22mm} : \text{UL problem} \\ \notag
    &s.t~  \text{Eqs.~}\eqref{eq:loss_revenue_robust} -\eqref{eq:premium_bound_robust} \\
    &\lambda_{t}^{c, WC}  \in \argmin\{ \\
    & \text{Eqs.~}\eqref{eq:EVCS_CVAR_primal_obj}-\eqref{eq:primal_cvar_constraint_last} \hspace{9mm} : \text{ML problem} \\
    & \text{Eqs.~}\eqref{eq:gen_const}-\eqref{eq:load_const} \hspace{8.5mm}: \text{LL primal constraints} \\
    & \text{Eqs.~}\eqref{eq:dll_constraint1}-\eqref{eq:dll_constraint3}\hspace{8.5mm} : \text{LL dual constraints} \\
    & O_{LL}^{s} = O_{DLL}^{s}, ~\forall s \in \mathcal{S} \hspace{2mm} : \text{LL strong duality}\}.
\end{align}
\end{subequations}
 We use the C\&CG algorithm to solve the TL optimization in Eq.~\eqref{eq:tri_robust} as applied in \cite{zeng2013solving,ruiz2015robust}. The C\&CG algorithm is based on the decomposition of the TL optimization in Eq.~\eqref{eq:tri_robust} into  the principal problem and subproblems: 

\subsubsection{Principal problem}
\begin{subequations}
\begin{align}
    \min_{\Xi^{MP}}~& x^{WC} + ||{\boldsymbol{\lambda^{c,WC}}}||_2^2 ; 
    \\
    &s.t.~ \text{Eqs.~}\eqref{eq:loss_revenue_robust}-\eqref{eq:premium_bound_robust} \hspace{2.5mm} \text{: UL constraints}\\
    & \text{Eqs.~}\eqref{eq:primal_cvar_constraint1} - \eqref{eq:primal_cvar_constraint_last}  \hspace{8.5mm} \text{: ML constraints} \\
    & \text{Eqs.~}\eqref{eq:gen_const}-\eqref{eq:load_const} \hspace{8.5mm} \text{: LL primal constraints} \\
&\text{Eq.~}\eqref{eq:dll_constraint1}-\eqref{eq:dll_constraint3}\hspace{9 mm} \text{: LL dual constraints} \\
& O_{LL}^{s} = O_{DLL}^{s}, ~\forall s \in \mathcal{S} \hspace{1mm} \text{: LL strong duality}\},
\end{align}
with the C\&CG constraints updated at every iteration $k$ as:

\begin{align}
\label{eq:CCG_1}
    &||{\boldsymbol{\lambda^{c,WC}}}||_2^2 \geq ||{\boldsymbol{\lambda^{c, WC,(j-1), *}}}||_2^2,  ~ j = 2, \dots,k, \\ \label{eq:CCG_2}
    & \lambda_t^{c,WC} \geq \lambda_t^{c, WC, (j-1),*}, ~ \forall t \in \mathcal{T}, ~ j = 2, \dots,k, 
\end{align}
\end{subequations}

\subsubsection{Subproblem}

\begin{subequations}
\begin{align}
\min_{\Xi^{SP}} & ~||{\boldsymbol{\lambda^{c,WC}}}||_2^2 \\
& s.t.~ v + \sum_s \phi^s \zeta_s \leq 0, \\
&\alpha \zeta^s \geq -\tilde C^{WC}(d^s, \lambda^c, x^{(k),*}) -v, ~ \forall s \in 1:S \\
& \zeta^s \geq 0, ~ \forall s \in 1:S \\
& \lambda_t^{c,WC} \geq 0, ~ \forall t \in 1:T\\
& \text{Eqs.~}\eqref{eq:gen_const}-\eqref{eq:load_const} \hspace{8.5mm}: \text{LL primal constraints} \\
&\text{Eqs.~}\eqref{eq:dll_constraint1}-\eqref{eq:dll_constraint3}\hspace{8.5mm} : \text{LL dual constraints} \\
& O_{LL}^{s} = O_{DLL}^{s}, ~\forall s \in \mathcal{S} \hspace{2mm} : \text{LL strong duality}\}.
\end{align}
\end{subequations}
where $\Xi^{MP} =\{\Xi^{I} \cup \Xi^{E} \cup \Xi^{G}\}$ and $\Xi^{SP} = \Xi^{E} \cup \Xi^{G}$.  The upper  and lower bounds, for the convergence of the C\&CG algorithm, are calculated as the values of $x^{WC} + ||{\boldsymbol{\lambda^{c,WC}}}||_2^2$ in the subproblems and the principal problem, respectively. For any iteration $k$, the value of $x^{WC,(k),*}$ obtained from the principal problem is passed to the subproblems, which returns the value of $\lambda_t^{WC, c*,j}, \forall j \leq k$. In the first iteration, the principal problem is solved without considering C\&CG constraints in Eqs.~\eqref{eq:CCG_1} and \eqref{eq:CCG_2}.  The principal problem and subproblems are iterated until  the lower and upper bounds converge within an user-defined  tolerance. The additional cut in Eq.~\eqref{eq:CCG_2} is introduced in the ML objective to boost the convergence of the C\&CG algorithm. 
 
\vspace{-2mm}
\section{Case Study}
\label{sec:case_study}
\begin{figure}
  \centering
\includegraphics[width=1\columnwidth, clip=true, trim= 10mm 120mm 20mm 10mm]{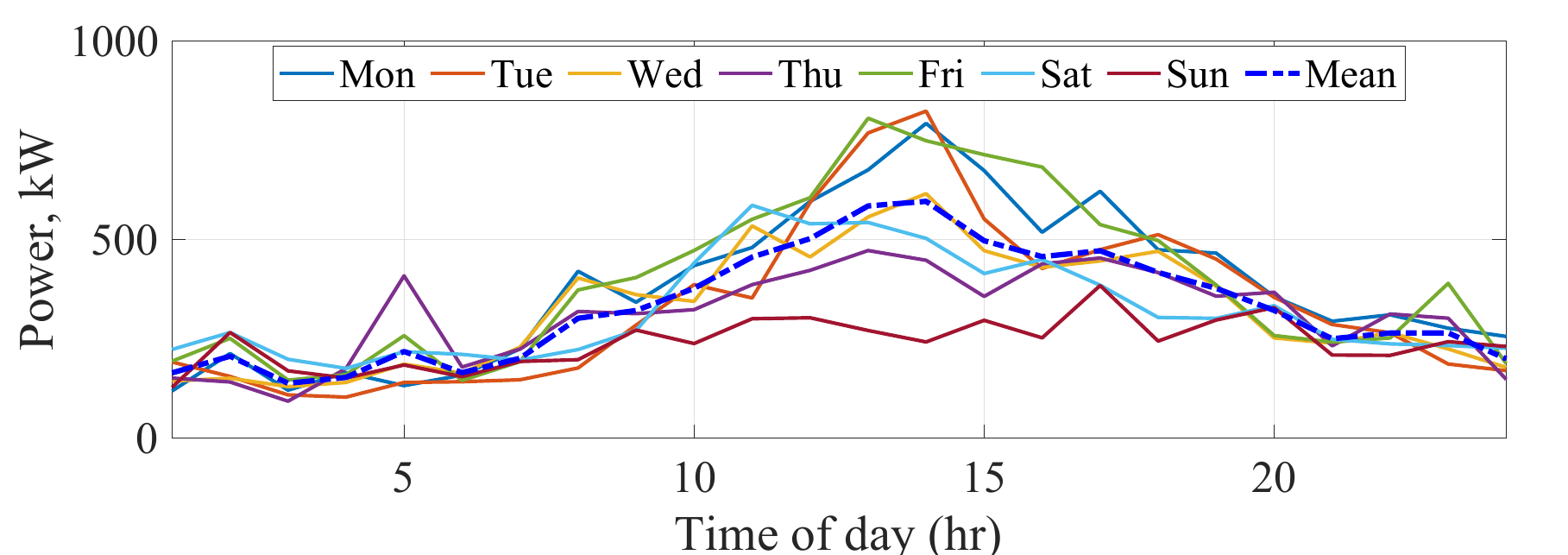}
 \caption{Hourly power consumption of EVCSs in Manhattan, NY.}
 \label{fig:scenario}
 \end{figure}

The case study is performed using  power grid and commercial EVCS data from Manhattan, NY. Fig.~\ref{fig:scenario} displays weekly  EVCS demand profiles collected as explained in \cite{acharya2020public}. Based on \cite{acharya2020public}, the power grid model is a 7-bus  network with 4 load buses (\#3, 4, 5, and 6) and 4 generation resources located at  buses  \#1, 2, 5, and 7. We use this system as a distribution network in this paper. The non-EVCS demand profiles at each bus  are built based on load distribution curves for New York City published by the New York Independent System Operator (NYISO).  This distribution network is connected to the NYISO system at buses \#1, 2, and 7, where we set their generation cost as the Locational Marginal Prices (LMPs) obtained from \cite{nyiso_LMP}. Additionally, generation resource at bus  \# 5 that is operated by the distribution system operator has an incremental production cost of \$10/MWh.

We define the data for the base rate-based insurance policy factors (denoted as $\Pi$ in Section~ \ref{sec:bilevel_equiv_insurance_premium}) to  customize the premium for EVCSs. The box-constrained confidence interval for the uncertain insurance policy factors are defined as $r =[0.25~ 0.35]$, $\kappa =[0.2~ 0.3]$, and $\mathbb{P}(A) =[0.03582~ 0.04378]$. The box limits for $r$ and $\kappa$ are constructed using industry practices across insurers \cite{romanosky2019content, fitch}, while the limits for $\mathbb{P}(A)$ are constructed around the value of $\mathbb{P}(A)$ obtained using the SMP in Section~\ref{sec:SMP}. The box limits of $\mathbb{P}(A)$ consider $\epsilon =\pm10\%$ in $\mathbb{P}(A)$ estimated by the SMP.
To compute $\mathbb{P}(A)$, Table~\ref{tab:SMP_parameters} shows the data-driven Weibull distribution parameters for the SMP state transitions and the data source for modeling the transition distributions. Table~\ref{tab:SMP_probability} shows the sojourn time and the steady-state probabilities of each SMP state.   Since there is limited data on EVCSs-related cyberattacks recorded publicly, we use cyberattack data from information technology industry available from SANS Institute \cite{sans}, the Canadian Institute for Cybersecurity \cite{CIC}, and empirical studies \cite{franke2014distribution, liu2019probabilistic}. The SANS Institute \cite{sans} provides global survey data on time required for state transitions I-D, D-C, and C-G in 2019. The Canadian Institute for Cybersecurity in \cite{CIC} provides data on time required to infiltrate its information technology test beds (state transition G-I) with different attacks, such as DoS and brute-force. We use this data to obtain  Weibull distribution parameters ($\alpha, \beta$). Liu \textit{et al.} in \cite{liu2019probabilistic} and Franke \textit{et al.} in \cite{franke2014distribution} provide values of the distribution parameters for  state transitions I-F and F-G, respectively.  

The case study is carried out on a 64-bit personal computer with a 2.2 GHz Intel Core i7 processor and 8 GB RAM. The implementation uses the Ipopt and Gurobi solvers. All the simulation instances are solved within 1 hour.

\begin{table}

\caption{Weibull distribution parameters for  SMP state transitions.}
\centering
\resizebox{0.85\columnwidth}{!}{

\begin{threeparttable}
    \begin{tabular}{c|c|c|c|c}
    \hline
         \multirow{2}{*}{State Transition}&\multicolumn{2}{c|}{Parameter}& \multirow{2}{*}{Data}&  \multirow{2}{*}{Source}\\
    \cline{2-3}&Shape $\beta$&Scale $\alpha$ & &\\
    \hline
    G-I &2.0675&18.8178& Real& \cite{CIC} \\
    I-D& 1.9293&16.0712& Real& \cite{sans}\\
    D-C&1.5698&18.4858& Real& \cite{sans}\\
    C-G& 1.3816&15.7033& Real&\cite{sans}\\
    I-F& 0.7000& 400.000& Approx.& \cite{liu2019probabilistic}\\
    F-G& 0.6783& 13.4487& Real& \cite{franke2014distribution} \\
    \hline
    \end{tabular}
    \begin{tablenotes}\footnotesize
\item Real data is from surveys and empirical studies. Approx. is approximated.
\end{tablenotes}
\end{threeparttable}
}
\label{tab:SMP_parameters}

\end{table}

\label{sec:results}

\begin{table}

\caption{Sojourn time and  steady state probabilities of  SMP states.}
\centering
\resizebox{0.77\columnwidth}{!}{

\begin{threeparttable}
    \begin{tabular}{c|c|c}
    \hline
         \multirow{2}{*}{State $s$}&\multicolumn{2}{c}{Entity}\\
    \cline{2-3} &Sojourn Time $T_s$ (hour) &Probability $\mathbb{P}_s$\\
    \hline
    G &9.1016&0.2010 \\
    I& 13.3431&0.2943\\
    D&11.7754&0.2366\\
    C&11.3659&0.2283\\
    F&19.8271&0.03980\\
    \hline
    \end{tabular}
\end{threeparttable}
}
\label{tab:SMP_probability}
\end{table}

\subsection{EVCS With A Predetermined Tariff}
\begin{figure}
  \centering
\includegraphics[width=1\columnwidth, clip=true, trim= 20mm 130mm 25mm 15mm]{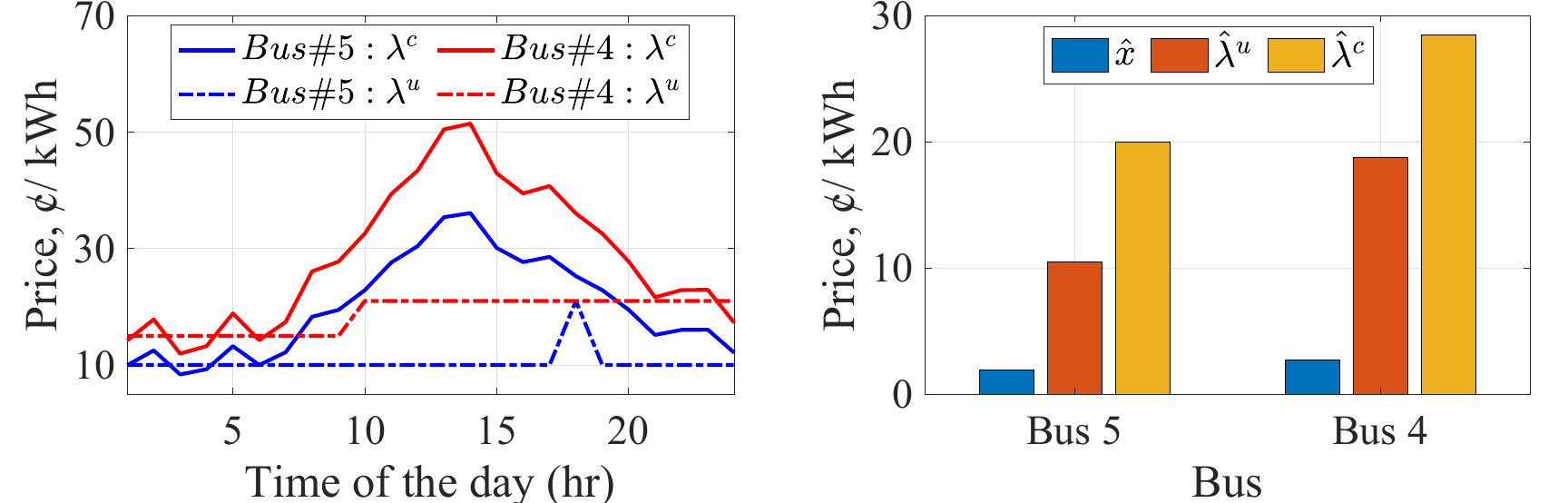}
 \caption{Performance of EVCS at buses $\#$4 and 5: (a) Hourly profiles of the predetermined electricity tariff ($\lambda^u$) and the EV charging price offered by the EVCS ($\lambda^c$) and (b) The insurance premium per kWh of the EVCS  demand ($\hat x$), the 24-hour average electricity tariff ($\hat \lambda^u$), and the 24-hour average EV charging price ($\hat \lambda^c$).  }
 \label{fig:sensitivity_LMP}
 \end{figure}

This section studies the performance of the  premium obtained using the analytical result in Theorem \ref{the:theorem1} and its robust and risk-averse reformulation described in Section~\ref{sec:robust}. Alongside, we study the sensitivity of the insurance premium to policy factors $\Pi$. For these analyses, we obtain the predetermined electricity tariffs in the power grid in two steps. First, the LMPs at buses \#1, 2, and 7, which are connected to the NYISO system, are obtained from the NYISO real-time dashboard  \cite{nyiso_LMP}. Second, the LMPs (``generation'' costs at buses \#1, 2, and 7) and the incremental production cost of the generator at bus \# 5 are fed to the optimization problem in  Eqs.~\eqref{eq:objective_lower_tri}-\eqref{eq:load_const} to determine the fixed tariff values. Note that we determine the value of the fixed tariff by solving Eqs.~\eqref{eq:objective_lower_tri}-\eqref{eq:load_const} independently from the tri-level model given in  Eqs.~\eqref{eq:UL-tri}-\eqref{eq:load_const}. The value of the predetermined electricity tariffs are shown in Fig.~\ref{fig:sensitivity_LMP}(a). Since the predetermined electricity tariff at bus \#5 is same as for other buses except bus \#4, we study these two buses. The predetermined electricity tariff $(\lambda^u)$ in Fig.~\ref{fig:sensitivity_LMP}(a) is independent of the EVCS demand, which follows from the dissimilarity of the hourly profiles for $\lambda^u$ and the EV charging price $(\lambda^c)$ in Fig.~\ref{fig:sensitivity_LMP}(a). Fig.~\ref{fig:sensitivity_LMP}(b) presents the  values of  $ \hat \lambda^u$, $\hat \lambda^c$, and $\hat x$ at buses \#4 and 5, where  $ \hat \lambda^u$ and $\hat \lambda^c$ are the average values of $\lambda^u$ and $\lambda^c$ in Fig.~\ref{fig:sensitivity_LMP}(a) and $\hat x$ is the insurance premium per KWh EVCS demand calculated using Eq.~\eqref{eq:x_analytical}. As explained in Section~\ref{sec:bilevel_equiv_insurance_premium}, $\sum_{s}\phi ^{s}\sum_{t}\hat xd_t^{s} = x$.
Since  $\hat \lambda^u$ at bus \#4 is greater than at  bus \#5, as Fig.~\ref{fig:sensitivity_LMP}(b) shows, $\hat \lambda^c$ and  $\hat x$ at bus \#4 are also greater than those  at bus \#5.

  \begin{figure}[!t]
  \centering
\includegraphics[width=1\columnwidth, clip=true, trim= 10mm 4mm 15mm 8mm]{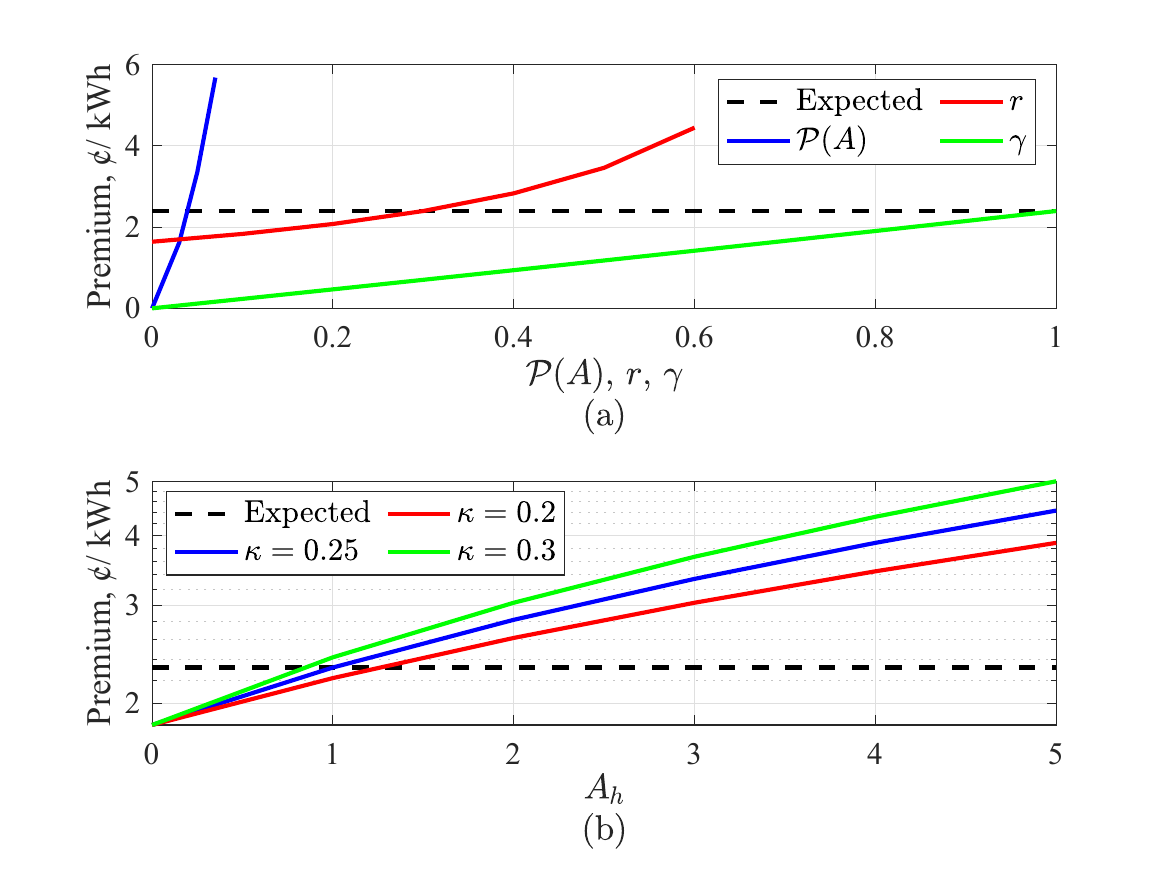}

 \caption{Sensitivity of the cyber insurance premium per kWh EVCS demand $(\hat x)$ based on the expected EVCS cost at bus \#4 with respect to changes in the policy parameters: (a) Changes in  $r$, $\mathbb{P}(A)$, and $\gamma$ and (b) Changes in $\kappa$ and $A_h$. The expected premium is calculated with $r=0.3$, $\mathbb{P}(A) = 0.03980$, $\gamma =1$, $\kappa = 0.25$, $A_h =1$, $\rho = 3$, and EVCS revenue $= \$ 908$/day. The daily average energy consumption and power rating of the EVCSs at bus \#4 are 3223 kWh and 2700 kW, respectively.}
 \label{fig:sensitivity_premium}

 \end{figure} 
 
Fig.~\ref{fig:sensitivity_premium} illustrates  the performance of the base rate premium as a function of policy factors $\Pi$ at bus \#4. To analyze the sensitivity of these factors, we study the analytically calculated expected insurance premium in Eq.~\eqref{eq:x_analytical}, which uses the mean value of $\mathbb{P}(A)$, $r$, and $\kappa$ obtained from their box-constrained confidence interval defined above, i.e., $\mathbb{P}(A) = 0.03989$, $r=0.3$, and $\kappa = 0.25$. Furthermore, we assume that the EVCS transfers $100\%$ of its cyber risk to the insurer, i.e., $\gamma =1$ and the EVCS has encountered one cyberattack in the past, i.e.,  $A_h =1$. In this setting, as Fig~\ref{fig:sensitivity_premium}(a) shows, the insurance design is more sensitive to $\mathbb{P}(A)$ than to other insurance policy factors. Similarly, the premium increases at a greater rate when the insurer is less confident about the values of the policy factors and the revenue reported by the EVCS, i.e., premium $x$ increases faster after the insurer increases profit loading factor $r$ above $50\%$ as seen in Fig.~\ref{fig:sensitivity_premium}(a). On the other hand, changes in risk transfer factor $\gamma$ affect the premium linearly. 

\begin{figure}[!b]
\centering
\includegraphics[width=1\columnwidth, clip=true, trim= 10mm 4mm 15mm 4mm]{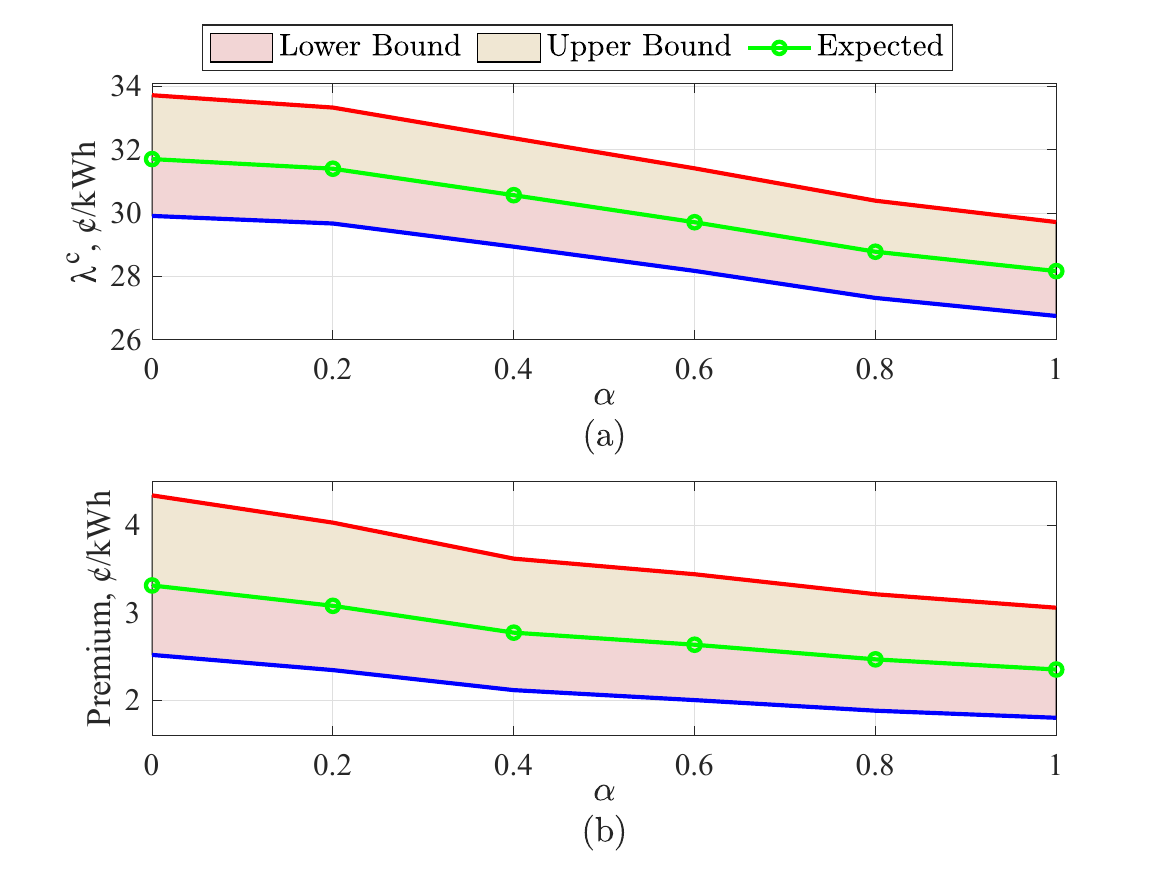}
 \caption{Changes in (a) the average EV charging price $(\hat \lambda^c)$ and (b) the cyber insurance premium per kWh EVCS power $(\hat x)$ for different levels of the EVCS risk-aversion measured by $\alpha$. The EV charging price and  premium  are bounded by the uncertainty in  policy parameters $\mathbb{P}(A)$, $r$, and $\kappa$.}
 \label{fig:bilevel_risk}
 \end{figure}
 
 Fig~\ref{fig:sensitivity_premium}(b) illustrates how the premium changes with the cyberattack history of the EVCS. A close-to-linear relationship is established between number of EVCS cyberattacks $A_h$ and premium $x$. An increase in $\kappa$ leads to a  faster premium increase. In addition to the premium obtained from expected cost in Fig.~\ref{fig:sensitivity_premium}, Fig.~\ref{fig:bilevel_risk} presents the bounds of premium and EV charging price for a risk-averse EVCS. The bounds are determined numerically using Eq.~\eqref{eq:bi-level-CVaR}. The lower and the upper bounds of $\hat x$ and $\lambda^c$ in Fig.~\ref{fig:bilevel_risk} use the lower and upper limits  of the box-constrained confidence interval for  policy parameters. That is, $\mathbb{P}(A) =0.03582$, $r =0.25$, and $\kappa = 0.2$ for lower bounds and $\mathbb{P}(A) =0.04378$, $r =0.35$, and $\kappa = 0.3$ for upper bounds. In this setting, as shown in Fig.~\ref{fig:bilevel_risk}(a), as the EVCS becomes more risk-averse (i.e., $\alpha$ decreases), it chooses the ``worse-case'' typical days. As a result, the EVCS optimization in Eq.~\eqref{eq:opt_EVCS_CVaR} chooses a greater value of $\lambda^c$ such that the EVCS net cost in Eq.~\eqref{eq:CVAR_EVCS_const} is at most zero. When $\lambda^c$ increases, as shown in Fig.~\ref{fig:bilevel_risk}(b), the optimization in Eq.~\eqref{eq:tri_robust} returns a higher value of $\hat x$. Moreover, as shown in Fig.~\ref{fig:bilevel_risk}, the bounds on $\hat x$ and $\lambda^c$ widen as the EVCS becomes more risk-averse (i.e., the value of user-defined parameter $\alpha$ reduces).  
 
\subsection{EVCS With A DLMP-based Tariff}
\begin{figure}
  \centering
\includegraphics[width=1\columnwidth, clip=true, trim= 20mm 118mm 30mm 0mm]{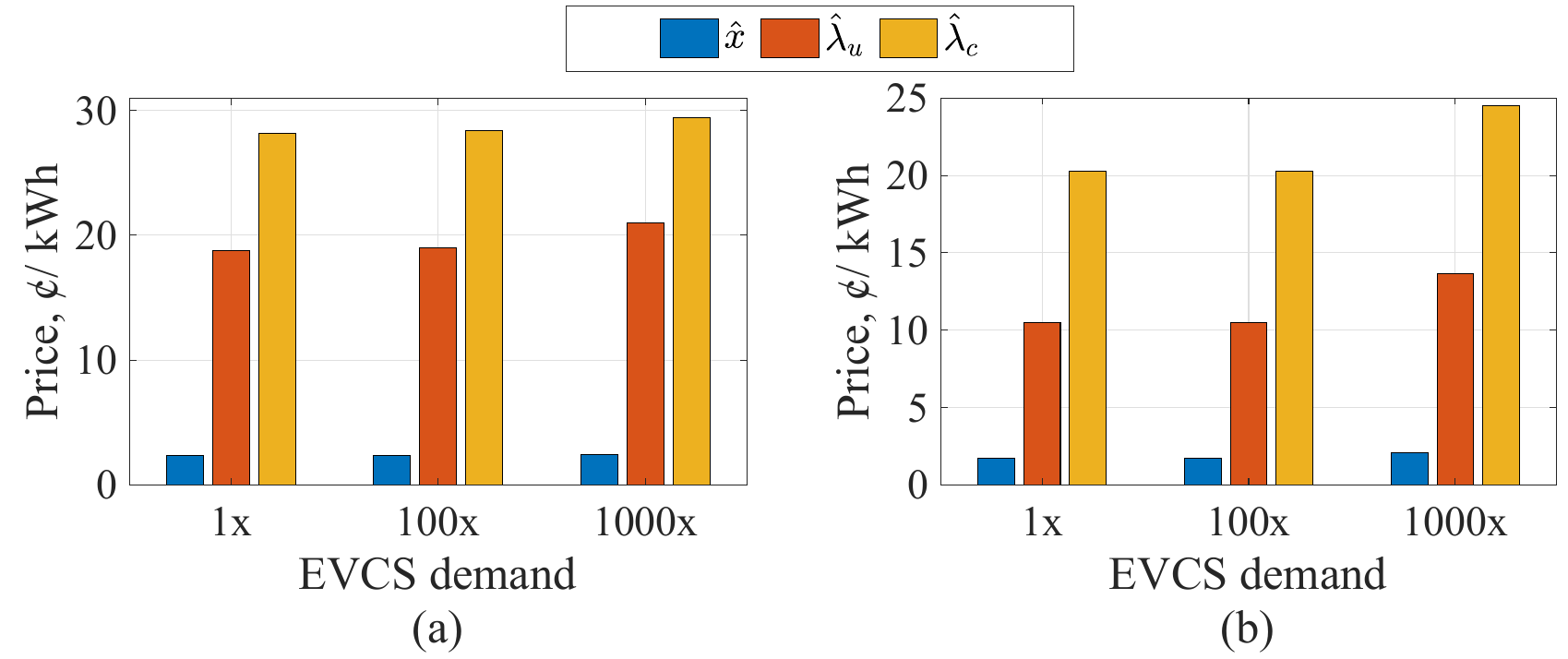}
 \caption{Changes in $\hat x$, $\hat \lambda^u$, and $\hat \lambda^c$ with an increase in EVCS demand by 100x and 1000x  at (a) bus \#4 and (b) bus \#5.}
 \label{fig:lmp_tri-level} 
 \end{figure}

 This section investigates the relationship between insurance premium $\hat x$ and dynamic electricity tariff $\lambda^{u}$ set by the utility using the DLMPs. Since the current penetration of EVCS demand as in Fig.~\ref{fig:scenario} is small relative to the system load, we analyze the effect of $\lambda^u$ on the premium design by increasing EVCS demands as indicated in Fig.~\ref{fig:lmp_tri-level} and Tables~\ref{tab:lambda_c_tri-level} and \ref{tab:premium_tri-level}.
 Fig.~\ref{fig:lmp_tri-level} present the sensitivity of the insurance premium, the EV charging price offered by the EVCS, and the electricity tariff at buses \#4 and  5 to an increase in the EVCS demand. As Fig.~\ref{fig:lmp_tri-level} shows $\hat x$, $\hat \lambda^c$, and $\hat \lambda^u$ start to change faster with a smaller increase in EVCS demand (e.g., 100x) at bus \#4 than at bus \#5. Thus, we focus on bus \#4. Table~\ref{tab:lambda_c_tri-level} presents the changes in the bounds of $ \hat \lambda^c$, designed by a risk-averse and robust EVCS, with an increase in the EVCS demand. As Table~\ref{tab:lambda_c_tri-level} shows the bounds on $\hat \lambda^c$ widen when the  EVCS becomes more risk-averse (i.e., the value of parameter $\alpha$ reduces). The result of Table~\ref{tab:lambda_c_tri-level} demonstrates that the EV charging price offered by the EVCS increases because of the increased EVCS demand in the distribution system, which is followed by the increase in the electricity tariff. Similarly, Table~ \ref{tab:premium_tri-level} shows the changes in the insurance premium with an increase in the EVCS demand. Similar to the $\hat \lambda^c$ in Table~\ref{tab:lambda_c_tri-level}, the premium in  Table~ \ref{tab:premium_tri-level} increases as the EVCS power capacity increases and the EVCS opts for a more risk-averse insurance design. 
 
 \begin{table}

\caption{$\hat \lambda^c$ with an increase in EVCS demand at bus \#4 (\textcent/\textrm{\normalfont kWh})}
\centering
\begin{threeparttable}
\resizebox{1\columnwidth}{!}{
    \begin{tabular}{|c|c|c|c|c|c|c|c|c|c|}
    \hline
       \multirow{2}{*}{d } &\multicolumn{3}{c|}{$\alpha = 1$}&\multicolumn{3}{c|}{$\alpha = 0.5$}&\multicolumn{3}{c|}{$\alpha = 0$} \\
        \cline{2-4} \cline{5-7} \cline{8-10}
     &LB &EX &UB &LB &EX &UB &LB &EX &UB \\
    \hline
     1$\times$&26.74&28.16&29.71&28.57&30.16&31.92&29.90&31.70&33.72\\ \hline
     100$\times$&26.95&28.37&29.93&28.80&30.40&32.16&30.18&31.98&34.01\\ \hline
     400$\times$&27.26&28.69&30.24&29.15&30.75&32.51&30.52&32.32&34.36\\ \hline
     800$\times$&27.68&29.11&30.67&29.69&31.29&33.07&31.18&32.99&35.04\\ \hline
     1000$\times$&27.98&29.41&30.98&30.08&31.69&33.47&31.66&33.49&35.55 \\ \hline
    \end{tabular}
    }
        \begin{tablenotes}\footnotesize
\item LB: Lower Bound, EX: Expected, UB: Upper Bound, d: EVCS Demand
\end{tablenotes}
\end{threeparttable}
\label{tab:lambda_c_tri-level}
\end{table}

 \begin{table}[!t]
\caption{$ \hat x$ with an increases in EVCS demand at bus \#4 (\textcent/\textrm{\normalfont kWh}) }
\centering
\begin{threeparttable}
\resizebox{1\columnwidth}{!}{
    \begin{tabular}{|c|c|c|c|c|c|c|c|c|c|}
    \hline
       \multirow{2}{*}{d } &\multicolumn{3}{c|}{$\alpha = 1$}&\multicolumn{3}{c|}{$\alpha = 0.5$}&\multicolumn{3}{c|}{$\alpha = 0$} \\
        \cline{2-4} \cline{5-7} \cline{8-10}
     &LB &EX &UB &LB &EX &UB &LB &EX &UB \\
    \hline
     1$\times$&1.801&2.351&3.057&2.062&2.699&3.519&2.519&3.310&4.338\\ \hline
     100$\times$&1.815&2.369&3.089&2.079&2.720&3.546&2.542&3.340&4.375\\ \hline
     400$\times$&1.836&2.395&3.111&2.105&2.754&3.588&2.570&3.375&4.420\\ \hline
    800$\times$&1.863&2.372&3.155&2.144&2.803&3.649&2.625&3.445&4.508\\ \hline
    1000$\times$ &1.884&2.409&3.187&2.172&2.838&3.694&2.666&3.497&4.573\\ \hline
    \end{tabular}
    }
    \begin{tablenotes}\footnotesize
\item LB: Lower Bound, EX: Expected, UB: Upper Bound, d: EVCS Demand
\end{tablenotes}
\end{threeparttable}
\label{tab:premium_tri-level}
\end{table}

\section{Discussion and Future Work}
The ratio of the cyber insurance premium to the revenue of the EVCS is calculated by dividing the values in Table~\ref{tab:premium_tri-level} by the corresponding values in Table~\ref{tab:lambda_c_tri-level}. This ratio is between 6.73$\%$--12.86$\%$, while the mean ratio is 9.35$\%$. These ratios are higher than those in other large-scale businesses (e.g., banking and financial services) because of two reasons \cite{deloitte}. First, as is customary in emerging insurance markets, we design the premiums for worst-case scenarios from the EVCS viewpoint. These scenarios include at least one attack on each given EVCS in the past ($A_h \geq 1$), full coverage of losses, high attack probabilities, high insurer profit loading factors, and risk-averse EVCS. Second, although  the cyber insurance premiums are becoming popular across the large businesses (they increased by $\approx 67\%$ in 2019--2020 \cite{deloitte}), they are nascent and unsuitable for comparison. For example, insurers are currently increasing their premiums to cover business losses caused by an improper assessment of cyber risk of the insured entities \cite{fitch}. To cover the cost of the cyber insurance designed in this paper, the risk-neutral and insured EVCS increases the EV charging price by $\approx 8\%$. The  premium  can  be  lowered,  if  the EVCS deploys defenses   reducing   its  attack probability as shown in Fig.~\ref{fig:sensitivity_premium}.

The insurance design is the first of its kind for high-wattage demand-side technologies (e.g., public EVCSs). Cyber risk assessment, the most critical aspect of the insurance design, can be extended in multiple directions. Cyber risk of a networked EVCS is correlated with its peers in the network. The accuracy of cyber risk assessment can be enhanced by considering such interdependence. Different attacks (e.g., phishing attack, DoS attack) can have different  motivations/objectives of the attacker and thus, these attacks could have different probabilities. Also, attackers can have a different time of interest, such as holidays, festivals, and peak-demand days. Such time-and attack-dependent probabilities could help  EVCSs to select proper defenses and reduce their premium.

\section{Conclusion}
 This paper develops models for optimizing cyber insurances  against cyberattacks on commercial EVCSs. These models internalize the operational dynamics of EVCSs and the power distribution company. The insurance premium for EVCSs under the predetermined electricity tariff is modeled as a bi-level optimization, while the premium for EVCSs operating under the DLMPs is modeled as a tri-level optimization. The paper makes the models robust to parameter uncertainty in insurance characteristics and risk attitudes of EVCS. The analytical results show that insurance design is sensitive to the probability of attacks on EVCSs in comparison to other design parameters. 
Numerical results show that premium and EV charging price increase for risk-aware EVCSs. 

\bibliographystyle{IEEEtran}
\bibliography{ref_insurance}
\end{document}